\documentclass[%
 aip,
 amsmath,amssymb,
 reprint,%
]{revtex4-2}

\usepackage{graphicx}
\usepackage{dcolumn}
\usepackage{bm}

\usepackage[utf8]{inputenc}
\usepackage[T1]{fontenc}
\usepackage{mathptmx}
\usepackage{etoolbox}

\makeatletter
\def\@email#1#2{%
 \endgroup
 \patchcmd{\titleblock@produce}
  {\frontmatter@RRAPformat}
  {\frontmatter@RRAPformat{\produce@RRAP{*#1\href{mailto:#2}{#2}}}\frontmatter@RRAPformat}
  {}{}
}%
\makeatother

\newcolumntype{M}{>{\vspace{1mm}\centering}m}
\makeatletter
\newcommand\newtag[1]{#1\def\@currentlabel{#1}\label{Tab1:#1}}
\makeatother
\newcommand{\n}{\newline}
\begin{document}

\title{A series of avoided
crossings of resonances in the system of several different
dielectric resonators results in giant $Q$-factors}

\author{K. Pichugin}
 \affiliation{Kirensky Institute of Physics, Federal Research Center
KSC SB RAS, 660036, Krasnoyarsk, Russia}
\author{A. Sadreev}
\email{almas@tnp.krasn.ru}
\author{E. Bulgakov}
\date{\today}
\begin{abstract}
We perform optimization of $Q$-factor in the system of
freestanding three/four/five/six coaxial subwavelength dielectric
disks over all scales. Each parameter contributes almost one order
of magnitude of the $Q$-factor due to multiple avoided crossings
of resonances to give totally the unprecedented values for the
$Q$-factors: $6.6\cdot10^4$ for the three, $4.8\cdot10^6$ for
four, $8.5\cdot10^7$ for five and one billion for six freestanding
silicon disks. By multipole analysis of the resulting hybridized
resonant mode we observe that such extremely large values of the
$Q$-factor are attributed to strong redistribution of radiation
that originates from almost exact destructive interference of
dominating complex multipole radiation amplitudes.
\end{abstract}

\maketitle

\section{Introduction}

Since the famous paper by Gustav Mie \cite{Mie1908}, the
engineering of dielectric cavities in optics and photonics has
been a long-standing area for the application of various ideas and
approaches to enhance the quality factor $Q$ due to its paramount
importance in both applied and fundamental research. However there
is a fundamental upper limit for $Q$-factor because of leakage of
radiation power from isolated dielectric resonator into the
radiation continuum \cite{Colton,Silveirinha14}. There are many
ways to enormously boost $Q$-factor. For example, one can use
Fabry-P\'{e}rot resonances or hide a cavity in photonic crystals
(PhC) \cite{Vahala2003,Ryckman2012,Seidler2013,Zhou2019}.
Whispering gallery modes (WGM) in the cavities with convex smooth
boundaries such as cylindrical, spherical or elliptical cavities
also show giant magnitudes of $Q$-factor
\cite{Braginsky1989,Gorodetsky2006,Acharyya2019,Jiang2020,Hong2022}.

Cardinally different way is bound states in the continuum (BICs),
which provide unique opportunity to confine and manipulate
electromagnetic wave within the radiation continuum (see reviews
\cite{Hsu2016,Azzam2020,Huang2020,Joseph2021,Koshelev2022,Hu2022}).
The phenomenon of BICs is based on that electromagnetic power can
leak only in selected directions given by diffraction orders, if
to arrange dielectric cavities into a periodical array
\cite{Hsu2013,Bulgakov2017,Koshelev2019}. Although in reality the
number of cavities $N$ in an array cannot be infinite, the
$Q$-factor grows fast with $N$: quadratically for symmetry
protected (SP) quasi-BICs
\cite{Taghizadeh2017,Bulgakov2017b,Sadrieva2019} or cubically for
accidental BICs \cite{Polishchuk,Bulgakov2017b,Sidorenko2021}.
However this way of engineering of quasi-BICs goes away the
dielectric structures (DS) from compactness. For example, to
achieve $Q$-factor of the order $10^5$, we need at least several
tens of silicon disks
\cite{Bulgakov2019a,Bulgakov2019b,Sadrieva2019} or silicon cuboids
\cite{Kornovan2021}. The best results for the $Q$-factor were
reported by Taghizadeh and Chung \cite{Taghizadeh2017} with
$Q\sim10^5$ for 10 long identical silicon rods. In general, all
the ways to achieve the extremely high $Q$-factor listed above
require an extended DS in which the mode volume grows too
\cite{Gao2019,Zhou2019}.

\begin{figure}[ht!]
\centering
\includegraphics[width=0.75\linewidth]{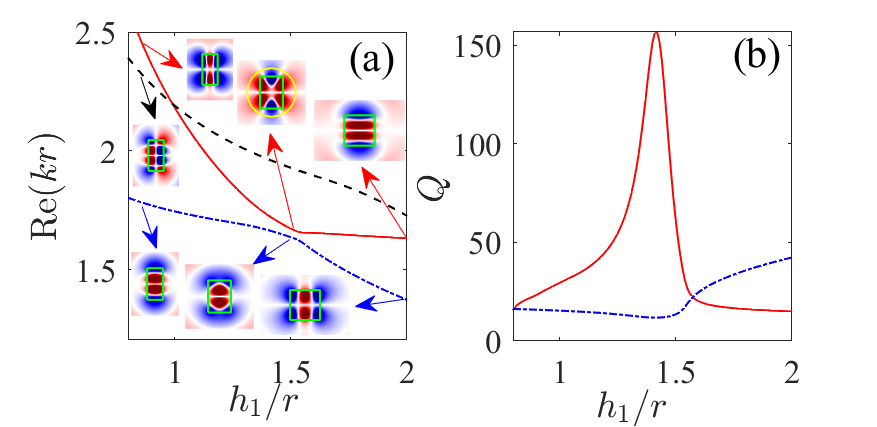}
\caption{(a) ACR of the two resonant even modes of a single
silicon disk with $\epsilon=12$ shown by red solid line and blue
dash-dot line for variation of aspect ratio \cite{Rybin2017}. The
insets show hybridization of modes (the tangential component of
electric field $E_{\phi}$ of TE modes). Black dash line shows the
evolution of the odd resonant mode, which is decoupled from former
modes. (b) Respective considerable enhancement of the $Q$-factor
due to ACR.} \label{fig1}
\end{figure}

In the present paper we show that this way of boosting the
$Q$-factor by a periodical array of $N$ identical resonators is
not optimal. We consider a system of only a few resonators each of
different scales variation of which causes a cascade of ACRs to
radically boost $Q$-factor. As a result, we have achieved giant
magnitudes of $Q$-factor considerably exceeding the results for
quasi-BICs maintaining nearly the same mode volume (see Table
\ref{Tab1}). ACR \cite{Brentano1990,Heiss2000} is a general and
fundamental phenomenon that describes the behavior of
eigenfrequencies of an open resonator, which are complex due to
coupling with the radiation continuum. Whether the resonant
frequencies exhibit either crossing or anticrossing depends on the
mechanism of interaction \cite{Heiss2000,Cao2015,Bernier2018}.
In any case, two resonances near ACR interfere in constructive and
destructive ways. The latter way enhances $Q$-factor and has been
successfully used in many types of single resonators
\cite{Wiersig2006,Yang2017,Huang2021}. As a result the Q-factor
can be strongly enhanced as has been demonstrated for different
choices of dielectric resonators
\cite{Wiersig2006,Song2010,Rybin2017,Koshelev2018,Chen2019,Wang2019,Odit2020,Volkovskaya2020,Huang2021}.
Example of such a super cavity mode due to hybridization of
resonances is highlighted by yellow open circle in Fig. \ref{fig1}
(a) for the case of disk shaped resonator. In a single silicon
disk with permittivity $\epsilon=12$ the $Q$-factor reaches
$Q=150$ for $h_1/r=1.4157$ as Fig. \ref{fig1} (b) shows.
Along with that the ACR approach to enhance the $Q$-factor has
been developed for a system of photonic molecules owing to
coupling between resonators
\cite{Boriskina2006,Boriskina2007,Song2010,Benyoucef2011,Chang2016,Pichugin2019,
Bulgakov2019,Dmitriev2019,Vinel2021,Hong2022}.


\section{The problem statement}

\begin{figure}[ht!]
\includegraphics[width=0.3\linewidth]{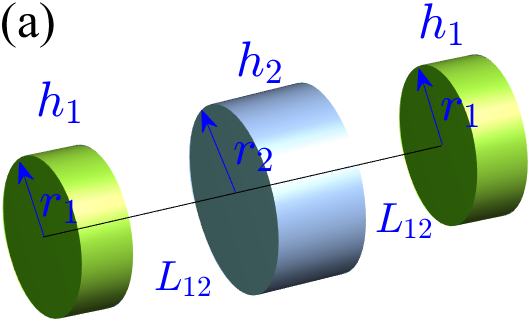}\hspace{4cm}
\includegraphics[width=0.35\linewidth]{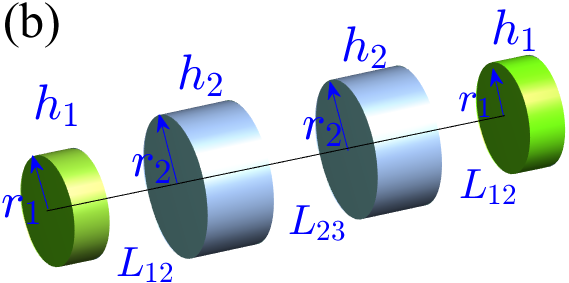}\\
\includegraphics[width=0.45\linewidth]{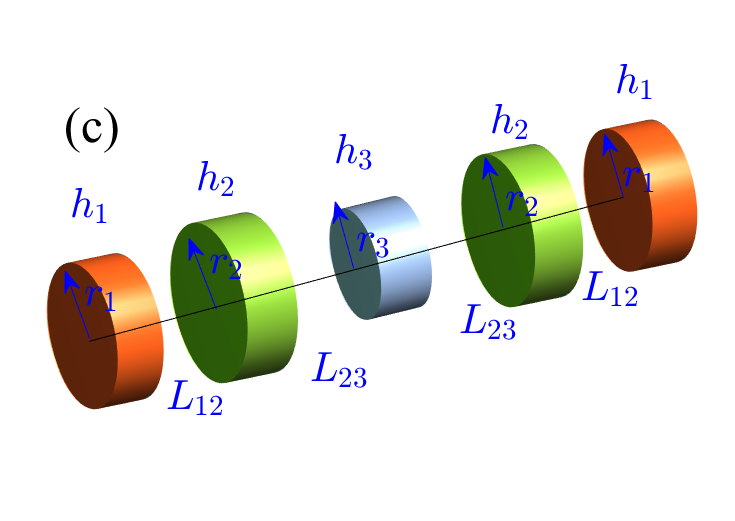}
\includegraphics[width=0.55\linewidth]{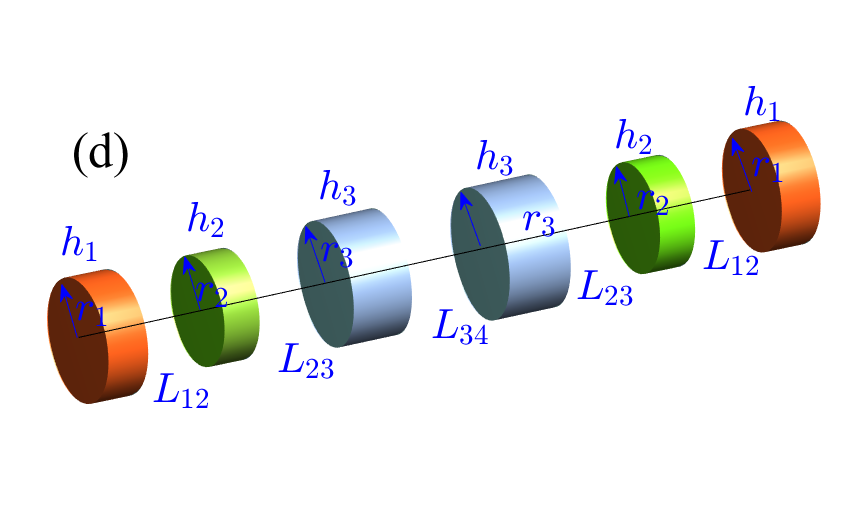}
\caption{(a) Disk inside dimer. (b) Dimer inside dimer. (c) System
(a) inside dimer. (d) System (b) inside dimer. All freestanding
disks with $\epsilon=12$ are coaxial but have different radii and
heights to form symmetric structure. Radius of the middle gray
disk(s) is used as scale $r$ in the systems. We assume azimuthal
number $m=0$ unless otherwise noted.} \label{fig2}
\end{figure}

Recently we have developed a way
to enhance the $Q$-factor by extending the number of resonators in
photonic molecule spaced at different distances on the example of
three and four coaxial silicon disks \cite{Pichugin2021}. As a
result we achieved $Q\approx10^6$ for four disks of identical
radii. In the present paper we put forward a novel strategy of
cascading ACRs in a system of $N$ disks in order to achieve
unprecedent magnitudes of $Q$-factor. We consider disks to be
freestanding and coaxial, made of silicon with permittivity
$\epsilon=12$ for wavelength $\lambda\approx1.55 \mu m$ at which
material losses are negligible \cite{Kitamura2007}. The eigenmodes
of the system are classified according to irreducible
one-dimensional representations of rotations around the symmetry
axis specified by the azimuthal index $m$. We will focus on the
case $m=0$ since the solutions of Maxwell equations are
additionally split by polarization that also simplifies the
problem. A system of $N$ coaxial disks offers $3N-1$ scales to
vary in general: $N$ radii $r_j$, $N$ heights $h_j, j=1, 2, \dots,
N$, and $N-1$ distances $L_{12}, L_{23}, \dots, L_{N-1 N}$.
Considering that one of the scales should be chosen for
dimensionless ratios, we obtain a total of $3N-2$ parameters.
Optimization over this number of parameters even for a small
number of resonators is an extremely time-consuming computational
problem. It is reasonable to choose systems that are symmetric
with respect to the inversion of the axis of rotational symmetry,
which radiates less compared to non-symmetric designs. Guiding by
this assumption, $N_p={\rm fix}((3N-1)/2)$ scale parameters are
left to vary, where a term 'fix' means a rounding to the nearest
integers towards zero. Particular cases of the systems with
$N=3,4,5,6$ are shown in Fig. \ref{fig2}.

Our central approach is based on dividing the system into two
subsystems: an internal subsystem of $N-2$ disks and an external
dimer represented by the first and the last disks. Assume, the
first inner subsystem has already been optimized to find the
hybridized resonant mode $\psi_{N-2}$ with maximal $Q$-factor.
This mode could be even or odd with respect to the axis inversion.
The outer dimer provides the resonant modes $\psi_2$ of the same
symmetry. For variation of scales of the dimer, we have multiple
ACRs of it's resonances with the optimized resonance of the
internal subsystem. As a result we obtain a hybridized resonant
mode $\psi_N$ with enhanced $Q$-factor of the total system.
However it must not be supposed that the solution of the problem
is finished. The interaction of two subsystems slightly perturbs
the optimized mode $\psi_{N-2}$, which obliges fine-tuning of the
inner subsystem. Therefore, we must continue the process of
successive optimizations.

Technically the strategy looks as follows. To enhance $Q$-factor
we perform an optimization procedure in parametric space for
initial sets of parameters. Each initial set lead to a local
maximum of the $Q$-factor. It is reasonable to fix at first the
scale parameters for inner subsystem of $N-2$ disks tuned to have
maximal $Q$-factor while the 3 remaining scales $L_{12}, r_1, h_1$
of outer dimer evolve in a three-dimensional parametric space. In
view of time consuming calculations we apply the Nelder-Mead
simplex optimization method for $Q$-factor in total
$N_p$-dimensional parametric space. As a result, we achieved giant
magnitudes of the $Q$-factor $6.6\cdot10^4$ for three,
$8.5\cdot10^7$ for five silicon disks at frequency $kr\approx1.75$
and $4.8\cdot10^6$ for four, $1.3\cdot10^9$ for six silicon disks
at frequency $kr\approx2.2$ maintaining nearly the same mode
volume (see Table \ref{Tab1}).

\section{A cascade of avoided crossings of resonances in the
system of several coaxial disks}\label{Sec3}

\subsection{Three disks}

\begin{figure}[ht!]
\centering
\includegraphics[width=0.75\linewidth]{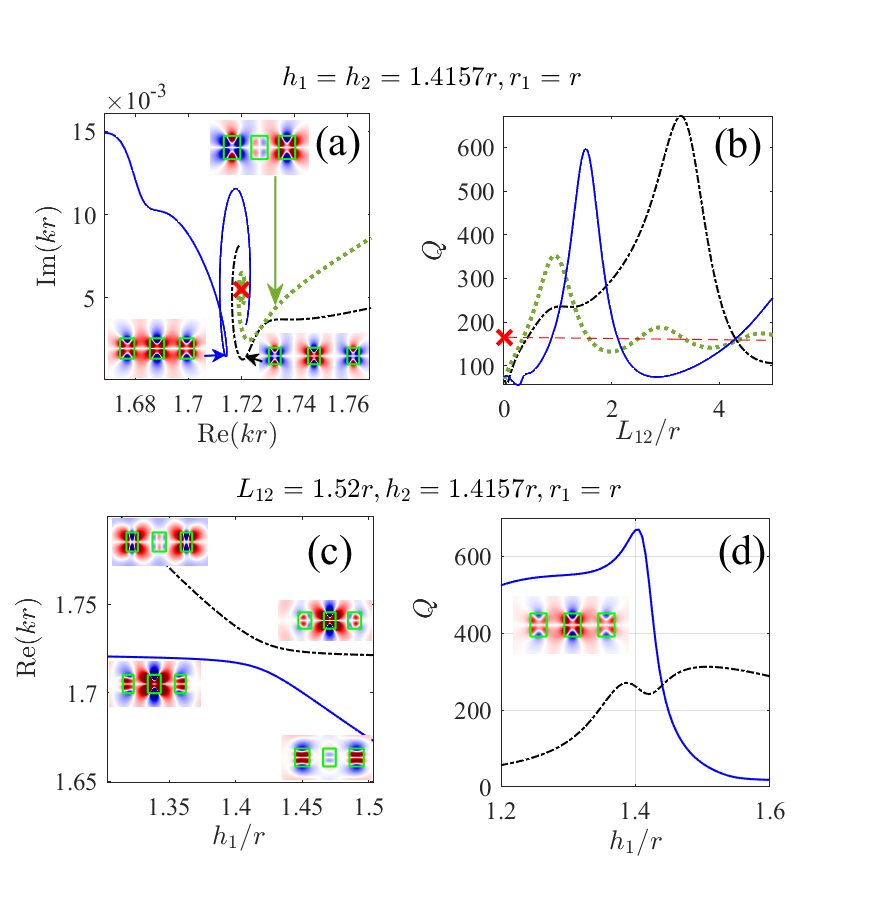}
\caption{(a) The first step of optimization over distance $L_{12}$
between 3 identical disks with fixed parameters $h_1=h_2=1.4153r$,
$r_1=r$. Corresponding dependence of the $Q$-factor (b). The
$Q$-factor reaches maxima at $L_{12}=0.945r, Q=350$ for green line
(mode profile ${\rm Re}(E_\phi)$ is shown on the middle inset of
(a)), $L_{12}=1.525r, Q=600$ for blue line and (mode profile is
shown on the left inset of (a)), $L_{12}=3.29r, Q=670$ for black
line (mode profile is shown on the right inset of (a)). (c) The
second step of optimization over $h_1/r$ with fixed
$L_{12}=1.52r$, $h_2=1.4157r$, $r_1=r$. The $Q$-factor (d) reaches
maximum at $h_1=1.4031r, Q=670$.} \label{fig3ad}
\end{figure}

To illustrate the strategy for achieving maximal $Q$-factor
outlined in section II we consider at first the system of 3 disks
sketched in Fig. \ref{fig2} (a) with 4 independent parameters
$r_1, L_{12}, h_1, h_2$ referred to the radius $r$ of central
disk.
First, we place already optimized one disk $h_2/r=1.4157$ (see
mode \ref{Tab1:1} in Table \ref{Tab1}) between 2 disks with the
same aspect ratio and vary distance $L_{12}$ only. At large
distance $L_{12}/r$ the resonances are almost degenerate and
marked by 'x' in Fig. \ref{fig3ad} (a). Drawing closer, the disks
interact according to the law $e^{ikL_{12}}/L_{12}^2$
\cite{Bulgakov2021} because of radiation of leaky resonant modes
by one disk and subsequent scattering by the others. Couplings
between these supercavity modes splits them into three modes with
spiral behavior shown in Fig. \ref{fig3ad} (a) with three
$Q$-factor peaks at corresponding distances as plotted in Fig.
\ref{fig3ad} (b). Insets show field configurations of disks
related to these $Q$-factor peaks. As a result, we obtain a total
gain in $Q$-factor 4 times more than for the case of the single
disk marked by red cross in Fig. \ref{fig3ad} (b).

In the next step of the optimization method, we fix
$L_{12}/r=1.52$ at which we get maximum value of $Q=600$ and vary
the height of outer disks $h_1$. This variation gives the ACR of
Fabry-P\'{e}rot-like mode, which depends strongly on $h_1$, and
Mie-like mode, which depends weakly on $h_1$. This phenomenon,
illustrated in Fig. \ref{fig3ad} (c), leads to a further
enhancement of the $Q$-factor up to $Q=670$ as shown in Fig.
\ref{fig3ad} (d).

\begin{figure}[ht!]
\centering
\includegraphics[width=0.65\linewidth]{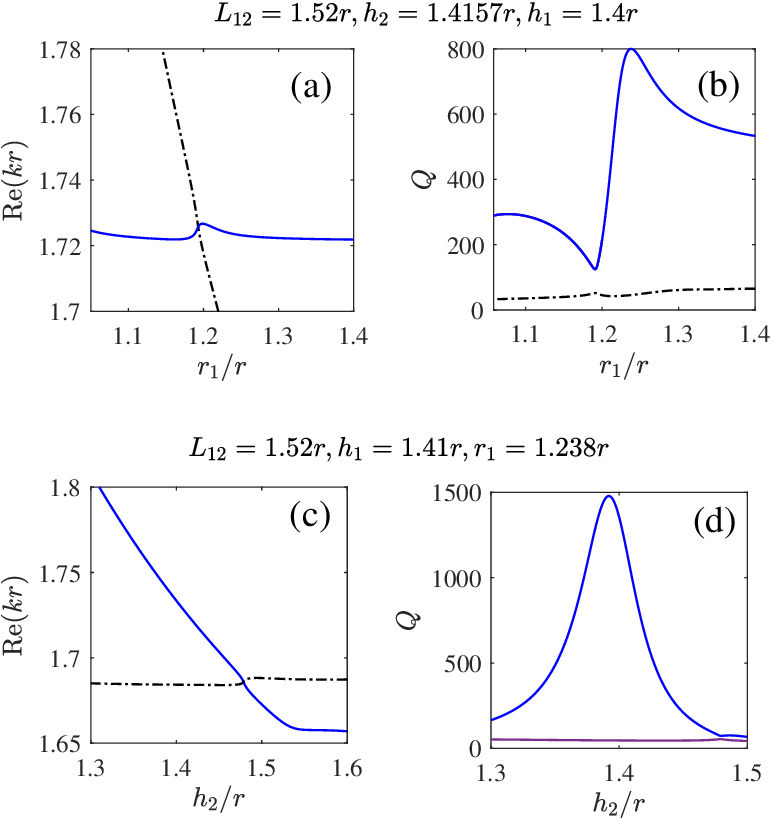}
\caption{Consequent ACRs over variation of different scales in the
system of three disks.} \label{fig3eh}
\end{figure}

At the third step, we allow the radius $r_1$ of the outer disks to
vary, while all other parameters are fixed to match the maximum
$Q$-factor in the second step. In contrast to the previous case,
the Mie-like resonant mode of outer disks is strongly dependent on
$r_1$, while the Fabry-P\'{e}rot-like mode is weakly dependent on
$r_1$. A selected event of ACRs of these modes is shown in Fig.
\ref{fig3eh} (a), which again raises the $Q$-factor up to 800.

Variation of height of the inner disk $h_2$ while all other
parameters are fixed for $Q=800$ closes the first round
optimization procedure in full four-dimensional parametric space.
The fourth step shown in Fig. \ref{fig3eh} (c) boosts the
$Q$-factor twice compared to the previous step, as seen from Fig.
\ref{fig3eh} (d). Repeating these rounds, at the end of the
optimization procedure we obtain $Q=6.6\cdot10^4$ (see the mode
\ref{Tab1:4} in Table \ref{Tab1}).

Another way to shed light on the enormous enhancement of the
$Q$-factor is to see the evolution of resonances along the
trajectory obtained by the traditional gradient descent method.
This method give us point ${\bf X}_{\infty}$ in a parametric space
with local maximum of $Q$-factor, which are the limit point of
iterations
\begin{equation}\label{GD}
{\bf X}_{n+1}={\bf X}_n+\eta\nabla Q({\bf X}_n), n\to\infty
\end{equation}
with appropriately chosen step $\eta$.

\begin{figure}[ht!]
\centering
\includegraphics[width=0.65\linewidth]{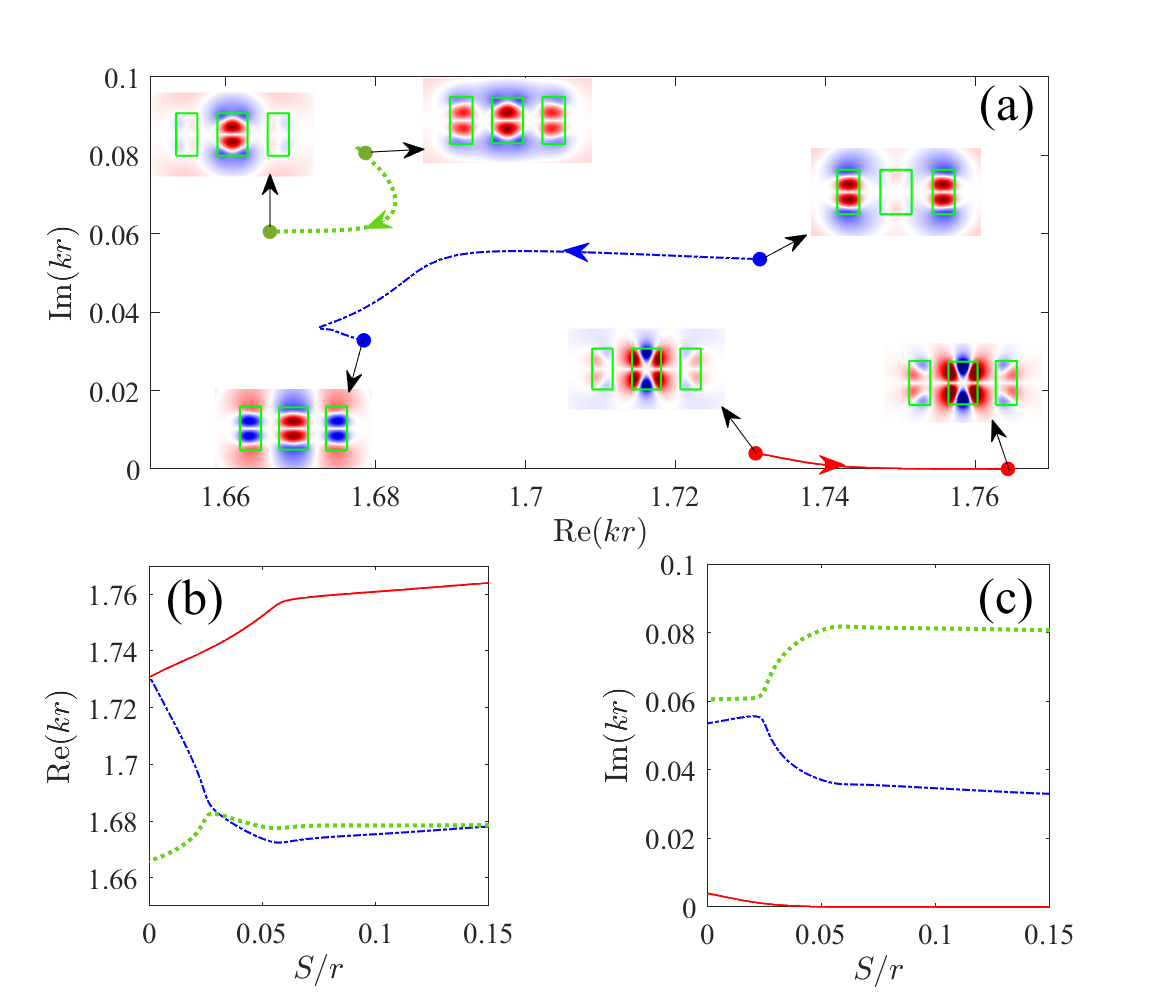}
\caption{(a) Evolution (\ref{GD}) of complex eigenfrequencies in
the four-dimensional parametric space of all scales in a system of
three coaxial disks. ${\bf X}_0$ is given by $r_1=r, h_1=r,
L_{12}=0.945r, h=1.4157r$. The final point ${\bf X}_\infty$
correspond to the mode \ref{Tab1:4} in Table \ref{Tab1}.
}\label{conv}
\end{figure}

The total result of the method can be represented as a trajectory
in the four-dimensional parametric space, whose length is
determined as a curvilinear integral
\begin{equation}\label{traj}
    S=\int_{{\bf X}_0}^{{\bf X}_\infty} \nabla F d{\bf s},
\end{equation}
where
\begin{equation}\label{F}
    F=\frac{1}{r}\sqrt{h_1^2+h_2^2+L_{12}^2+r_1^2}
\end{equation}
and ${\bf X}_0$ is an initial point of evolution. The evolution of
the three relevant complex eigenfrequencies of the three-disk
system is presented in Fig. \ref{conv} as a function of length
$S$. In Fig. \ref{conv} we can see typical ACR-like behavior: the
real parts of the green and blue lines cross (Fig. \ref{conv} (b))
while the imaginary parts repel each other (Fig. \ref{conv} (c)).
The interaction of at least 3 eigenfrequencies results in enormous
enhancement of the $Q$-factor on the red line.

There are a few points in the full four-dimensional parametric
space to which the optimization method converges. In this section,
we present only the most outstanding results, which exceeds values
of the first iteration shown in Fig. \ref{fig3eh} by several
orders of magnitude.
\begin{figure}[ht!]
\centering
\includegraphics[width=0.65\linewidth]{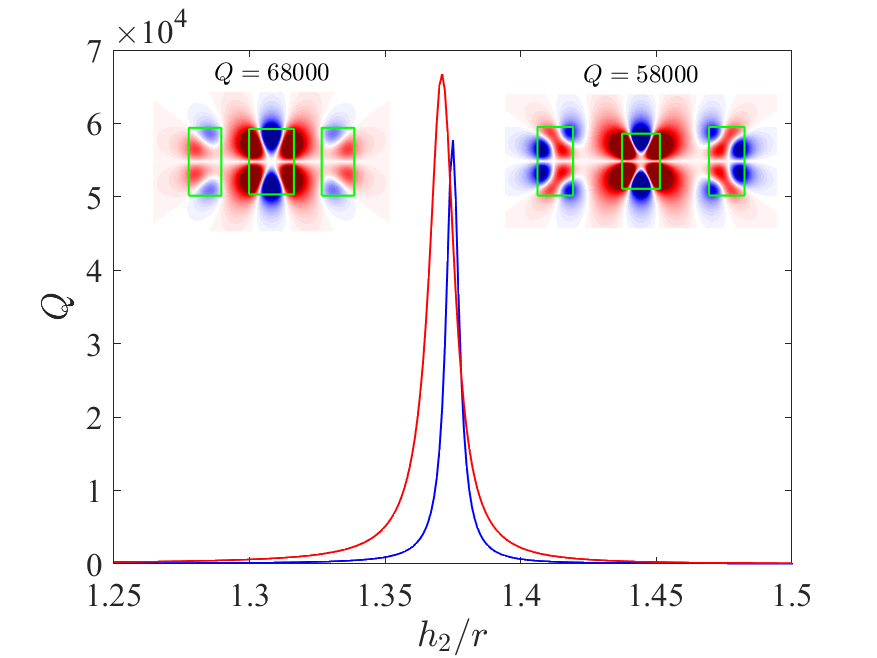}
\caption{Dependence of the $Q$-factor on $h_2/r$ for optimized
modes \ref{Tab1:3} (blue) and \ref{Tab1:4} (red) in Table
\ref{Tab1}. }\label{fig4}
\end{figure}
Among the local maxima of the $Q$-factor there are
eigenfrequencies with unprecedented $Q=5.8\cdot10^4$ and
$Q=6.6\cdot10^4$, as shown in Fig. \ref{fig4}. In both cases, we
have very similar supercavity modes in the middle disk while the
structure of EM field in the outer dimer is different. It is worth
noting that optimized mode for 5 disks (mode \ref{Tab1:5} in Table
\ref{Tab1}) is somewhat combination of the above modes: EM field
in disk 1 looks like EM field in disk 1 of mode \ref{Tab1:4} while
EM field in disk 2 looks like EM field in disk 1 of mode
\ref{Tab1:3} in Table \ref{Tab1}.
Thus, we can conclude that the outer dimer plays an important role
in resonant shielding of the supercavity mode radiation. Because
of the even symmetry of the supercavity mode, there are no ACR of
this mode with resonant odd dimer modes.

Next, let us consider the four coaxial disks sketched in Fig.
\ref{fig2} (b). By virtue of inversion symmetry, we represent the
system in the form of two dimers: internal and external. In total,
we have five scale parameters for ACRs: two heights $h_1$ and
$h_2$,
the two lengths of the dimers expressed via the two distances
$L_{12}$ and $L_{23}$, and, finally, the radius $r_1$ of the outer
dimer. All parameters are considered in respect to the radius of
the inner dimer $r$.
\begin{figure}[ht!]
\centering
\includegraphics[width=0.65\linewidth]{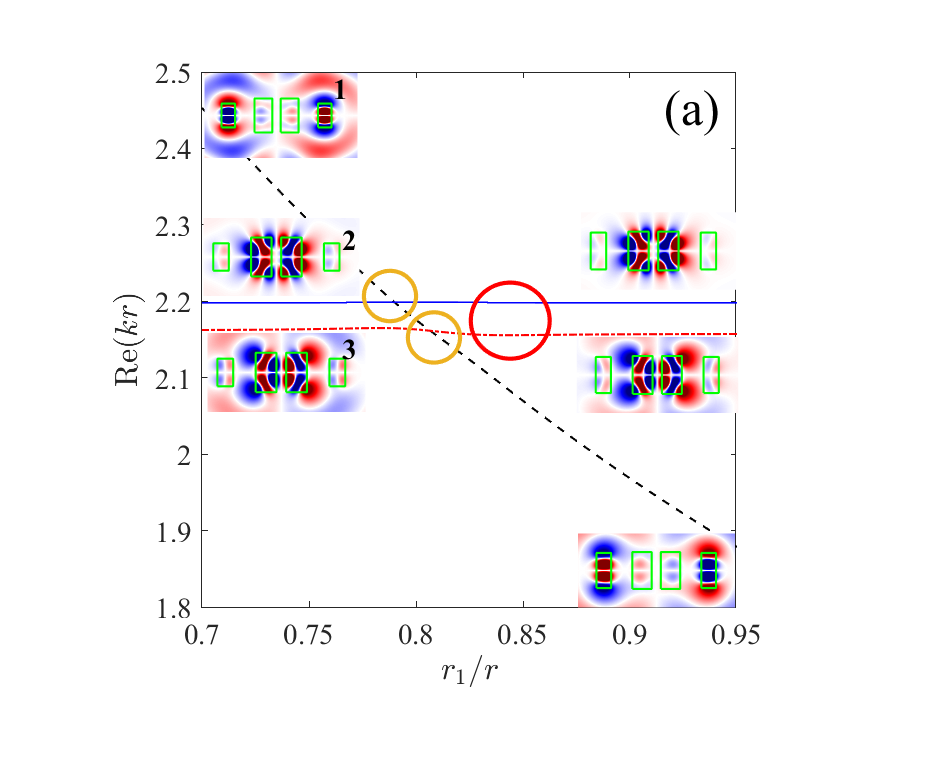}
\includegraphics[width=0.65\linewidth]{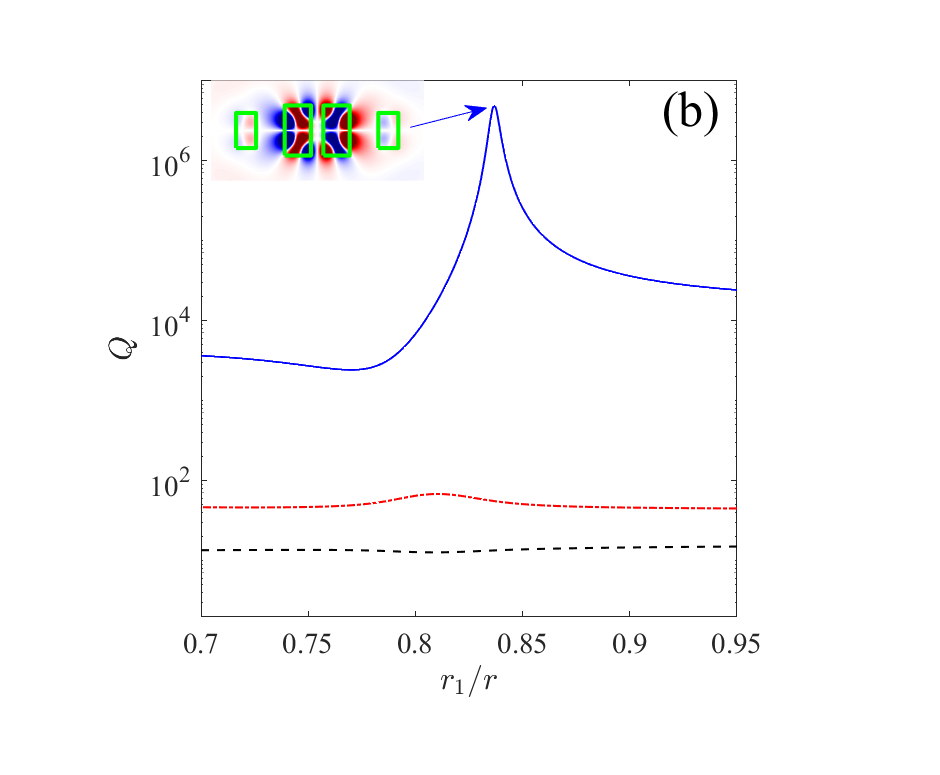}
\caption{(a) The final step of ACR for variation of radius $r_1$
of external dimer relative to the radius of internal dimer $r$
with strong enhancement of the $Q$-factor (b).} \label{fig5}
\end{figure}
We omit the iteration steps for all five parameters of the system
of two dimers. In addition, the reader can find some scenarios for
ACRs for variation of four scales: two heights and two distances
for identical radii $r_1=r$ in our previous publication
\cite{Pichugin2021}. This allows to boost the $Q$-factor up to one
million (see mode \ref{Tab1:7} in Table \ref{Tab1}). In the
present paper, we perform the final step by optimizing all 5
parameters.

Similar to the case of three disks, shown in Fig. \ref{fig3ad}, we
have a Mie-like mode of the outer dimer, labelled as 1 in Fig.
\ref{fig5} (a), which strongly depend on the radius $r_1$ of
external dimer. The other two Fabry-P\'{e}rot-like modes, labelled
as 2 and 3, are mostly localized in the inner dimer have weak
dependence on $r_1$. As a result, we observe a cascade of ACRs
around $r_1/r=0.8$, highlighted by yellow open circles, which,
however, do not lead to magnificent enhancement of $Q$-factor. In
contrast to these conventional ACRs, we observe a slightly
noticeable ACR around $r_1/r=0.85$ with
the Fabry-P\'{e}rot-type-like modes of outer dimer. This results
in a giant boosting of $Q$-factor up to almost five millions, and
can be explained by a cumulative effect of the interaction of
nearby resonances.


In Table \ref{Tab1} we collect the final configurations of systems
of five and six freestanding coaxial silicon disks after the
optimization procedure in parametric spaces of dimensions 7 and 8,
respectively. It can be seen that the outer dimer almost
completely shields radiation from the inner subsystem due to the
ACR of Fabry-P\'{e}rot-like resonant mode of external dimer with
resonant mode of internal subsystem, which has been already
optimized for the maximum $Q$-factor. These cases show
more impressive $Q$-factor results about $5.8\cdot10^7$ and one
billion (see Table \ref{Tab1}). Similar results can be achieved by
employing WGM modes in resonators or periodical quasi-BICs,
however, at the cost of increasing the radius or number of
identical resonators. This, in turn, increases mode volume.

\section{Multipole radiation for avoided crossing of resonances}

There is a useful tool to understand the nature of the extremely
high quality factor through multipole decompositions
\cite{Jackson}. This tool sheds light on the origin of high
$Q$-factor in the isolated disk \cite{Chen2019,Bogdanov2019} and
the origin of bound states in the continuum
\cite{Sadrieva2019a,Pichugin2021}. In far field region, EM field
can be expanded as
\begin{equation}\label{MN}
\mathbf{E}(\mathbf{x})=\sum_{l=1}^{\infty}\sum_{m=-l}^l[a_{lm}\mathbf{M}_{lm}(\mathbf{x})+
b_{lm}\mathbf{N}_{lm}(\mathbf{x})],
\end{equation}
where $\overrightarrow{M}_l^m$ and
$\overrightarrow{N}_l^m=\frac{1}{k}\nabla\times\overrightarrow{M}_l^m$
are the vector spherical harmonics \cite{Stratton,Linton2013}.
Then the relative radiated power of each electric and magnetic
multipole of order $l$ is
determined by the squares of the decomposition amplitudes
\cite{Jackson}
\begin{equation}\label{Plm}
    P_{l0}=P_{l0}^{TE}+P_{l0}^{TM}=P_0^{-1}[|a_{l0}|^2+|b_{l0}|^2]
\end{equation}
where $P_0$ is the total power radiating through the sphere with
large radius $ P_0={\sum_{l=1}^{\infty}|a_{l0}|^2]+|b_{l0}|^2]}$.
For the present case of coaxial disks with inversion symmetry and
azimuthal number $m=0$, the decomposition (\ref{MN}) is
substantially reduced to have an even $l$ for the symmetric
solutions shown in Fig. \ref{fig4}, and odd $l$ for the
antisymmetric solutions shown in Fig. \ref{fig5}
\cite{Bulgakov2021b}.

\begin{figure}[ht!]
\centering
\includegraphics[width=0.75\linewidth]{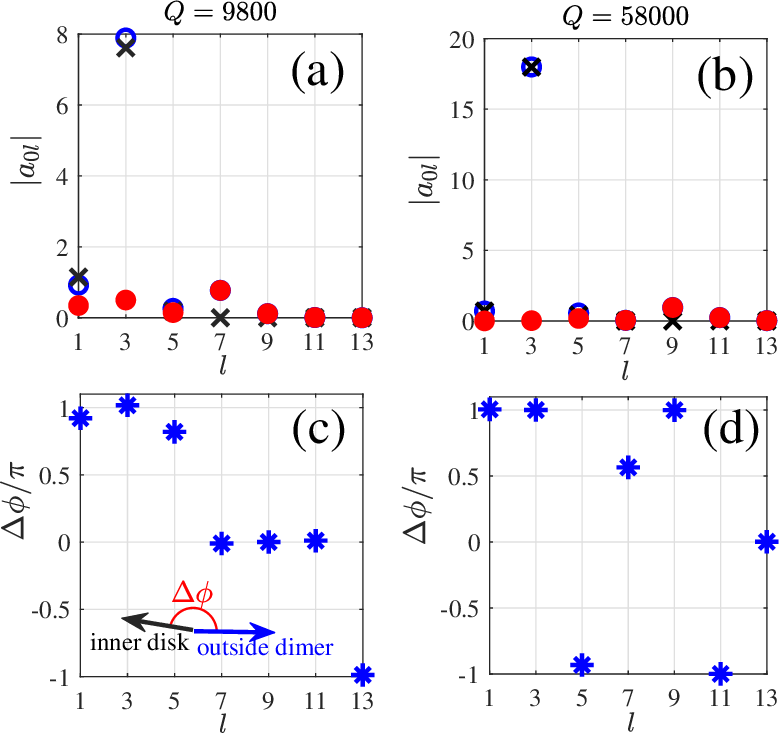}
\caption{The multipole radiation amplitudes $a_{l0},
l=1,3,5,\dots$ in Eq. (\ref{MN}) from the system of three disks of
identical radii
(mode \ref{Tab1:2} in Table \ref{Tab1}, $Q=9.8\cdot10^3$) (a), (c)
and of different radii (b), (d)
(mode \ref{Tab1:3} in Table \ref{Tab1}, $Q=5.8\cdot10^4$). The
amplitudes of inner/outer dimer disk marked by crosses/open
circles and amplitudes of the total system of three disks are
marked by red closed circles.} \label{fig6}
\end{figure}

The extreme $Q$-factor is associated with a strong redistribution
of multipole radiation towards high-order multipoles, because of
almost exact total destructive interference of low-order multipole
amplitudes. Using the formalism described in Ref. \cite{Doicu2006}
(Eq. (1.69)), we separate contributions from subsystems assembling
the DS in the far field region. For the case of three disks, we
distinguish multipole radiation from the inner disk and outer
dimer, whose complex amplitudes $a_{l0}$ in the series Eq.
(\ref{MN}) are presented in Fig. \ref{fig6}.
On subplots (a) and (b), the markers 'o' and 'x' correspond to
amplitudes $|a_{l0}|$ of the multipole radiation from the
subsystems of the inner disk and outer dimer, respectively, while
the red closed circles show the multipole coefficients of the
total DS, normalized by $P_0=\sum_l |a_{l0}|^2=1$. Subplots (c)
and (d) show the phase difference between the complex amplitudes
of the multipole radiation of the subsystems: the inner disk and
outer dimer. Left panels of Fig. \ref{fig6} show the case of
maximum $Q$-factor $9.8\cdot10^3$ achieved by ACR when optimizing
3 parameters $h_1$, $h_2$ and $L_{12}$ with $r_1=r$.
One can see strong multipole radiation for $l=3$ from both inner
disk and outer disks.
%
However, these complex amplitudes $a_{30}$ from both part,
sketched by arrows in the complex plane, have almost the same
moduli $|a_{30}|$ and a phase difference close to $180^\circ$,
which results in nearly full destructive interference of multipole
radiation at $l=3$. The total multipole radiation of the DS for
$l=3$ vanishes as shown by red closed circle in Fig. \ref{fig6}
(a). Note that there is still small multipole radiation at $l=7$
from the outer dimer, while the radiation from the inner disk is
mostly suppressed.

\begin{figure}[ht!]
\centering
\includegraphics[width=0.75\linewidth]{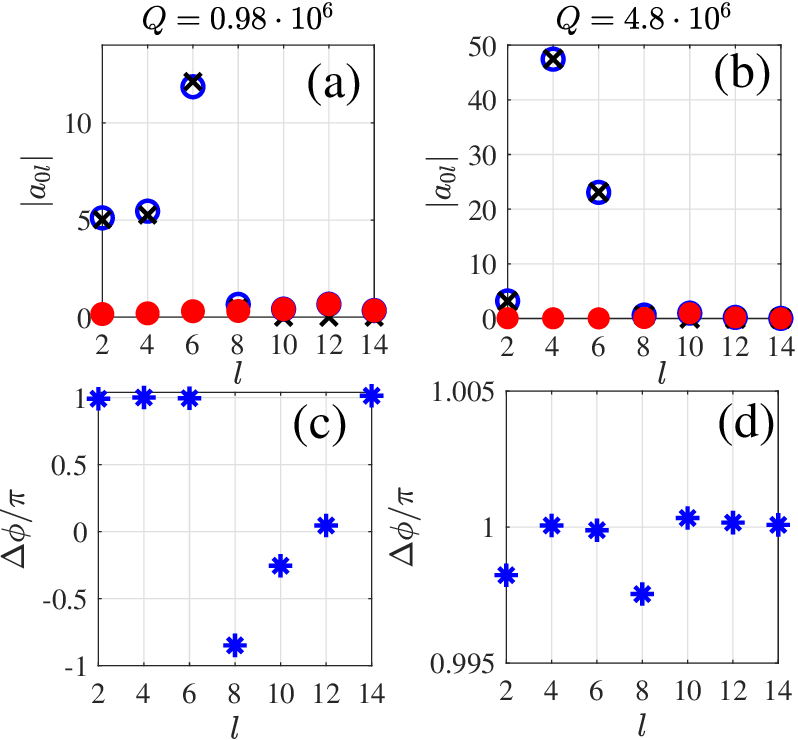}
\caption{The multipole radiation amplitudes $a_{l0},
l=2,4,6,\dots$ in Eq. (\ref{Plm}) for the system of four disks
with identical radii (mode \ref{Tab1:7} in Table \ref{Tab1},
$Q=9.8\cdot10^5$) (a) and (c).
(b) and (d) is amplitudes $a_{l0}$ for the system of four disks
with different radii
(mode \ref{Tab1:8} in Table \ref{Tab1}, $Q=4.8\cdot10^6$). The
amplitudes of inner/outer dimer of disks are marked by
crosses/open circles and amplitudes of the total system are marked
by red closed circles.} \label{fig7}
\end{figure}

Now let us consider the right panels of Fig. \ref{fig6}, which
show multipole radiation with $Q=5.8\cdot10^4$ achieved by
optimizing all possible parameters $h_1$, $r_1$, $h_2$ and
$L_{12}$. We can see from Fig. \ref{fig6} (b) that the additional
optimization over $r_1$ shifts the channel of maximum radiation
from $l=7$ to $l=9$ and increases the $Q$-factor by six times. One
can speculate that the introduction of an additional parameter to
vary
could suppress more contributions into multipole radiation. For
example, two dimers, as shown in Fig. \ref{fig2} (b), provide
more geometrical parameters than the previous case presented in
Fig. \ref{fig2} (a). The amplitudes of the multipole decomposition
are shown in Fig. \ref{fig7}. We can see remarkable effect of
destructive interference of complex multipole amplitudes $a_{l0}$.
The left panels demonstrate the effect till $l=4$, while the right
panels do it till $l=14$.

In subplots (c) and (d) the relative phases between amplitudes of
both dimers are shown. One can observe in Fig. \ref{fig7} almost
full destructive interference of the multipolar amplitudes at the
dominant channels $l=2,4,6$ from both dimers, when moduli of the
coefficients are equal, while phases differ by $\pi$. The
destructive interference of several amplitudes $|a_{0l}|$
simultaneously was achieved owing to the multiscale optimization
procedure only.

\section{Mode volumes of resonances with extremal $Q$-factor}

Optical cavities are able to trap light at discrete resonant
frequencies in a tiny volume in which the interaction of light
with matter can be dramatically enhanced via temporal and spatial
confinement of light. It is important not only to enhance the
$Q$-factor. Miniaturization of cavities with a high $Q/V_m$-ratio
is in demand to improve the light-matter interaction and reduce
layout for compact integrated optical circuits.

The $Q$-factor and effective mode volume $V_m$ -- two figures of
merit of optical cavities -- are of great importance in the
enhancement of light-matter interaction. The mode volume of a
dielectric cavity is given by the ratio of the total electric
energy to the maximum electric energy density \cite{Hu2016}
\begin{equation}\label{VM}
V_m=\frac{\int \epsilon({\bf x}) |{\bf E}({\bf x})|^2 dV}{{\rm
max}[\epsilon({\bf x}) |{\bf E}({\bf x})|^2]}.
\end{equation}
A summary of the $Q$-factors and mode volumes are collected by
Vahala \cite{Vahala2003} and range from $Q=2\cdot10^3,
V_m=5(\lambda)^3$ (FPR),  $Q=1.2\cdot10^4, V_m=6(\lambda)^3$ (WGM)
till $Q=1.3\cdot10^4, V_m=1.2(\lambda)^3$ (PhC cavity). Ultra-low
mode volumes in one-dimensional slotted photonic crystal single
silicon nanobeam cavities of order $(0.1-0.01)(\lambda/n)^3$ ($n$
is refractive index of DS) were reported
\cite{Ryckman2012,Seidler2013,Zhou2019}, however, at the cost of
compactness of resonator.
\begin{table*}
\begin{tabular}{|c|M{44mm}|m{6cm}|c|c|c|}
\hline
& Mode profile ${\rm Re}(E_\phi)$ & \hfil Scales\hfil & $\rm{Re}(kr)$ & $Q$ & $V_m\left(\frac{n}{\lambda}\right)^3_{\phantom{3}}$ \\
\hline \hline
\newtag{1}&\includegraphics[height=1.2cm,clip=]{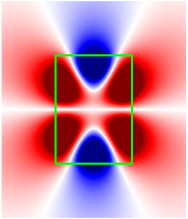} & $h_1/r=1.4157$ & 1.72 & $1.5\cdot10^2$ & 1.4\\
\hline
\newtag{2}&\includegraphics[height=1.2cm,clip=]{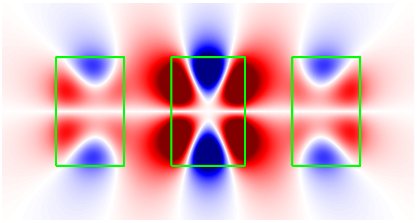} & $h_1/r=1.257, r_1/r=1$\n$h_2/r=1.362, L_{12}/r=0.873$ & 1.76 & $9.8\cdot10^3$ & 1.6 \\
\hline
\newtag{3}&\includegraphics[height=1.2cm,clip=]{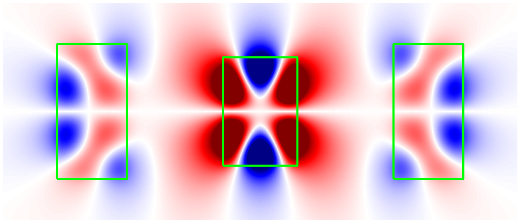} & $h_1/r=1.292, r_1/r=1.243$\n$h_2/r=1.375, L_{12}/r=1.78$ & 1.75 & $5.8\cdot10^4$ & 1.6 \\
\hline
\newtag{4}&\includegraphics[height=1.2cm,clip=]{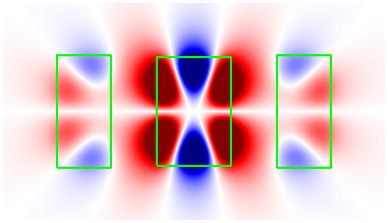} & $h_1/r=0.9972, r_1/r=1.0363$\n$h_2/r=1.3709, L_{12}/r=0.8497$ & 1.76 & $6.6\cdot10^4$ & 1.4 \\
\hline
\newtag{5}&\includegraphics[height=1.2cm,clip=]{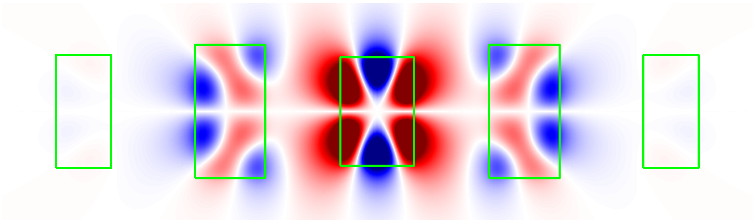} & $h_1/r=1.0237, r_1/r=1.0398$\n$h_2/r=1.3025, r_2/r=1.2319$\n$h_3/r=1.3629, L_{12}/r=1.5468$\n$L_{23}/r=1.3879$ & 1.77 & $8.5\cdot10^7$ & 1.7 \\
\hline \hline
\newtag{6}&\includegraphics[height=1.2cm,clip=]{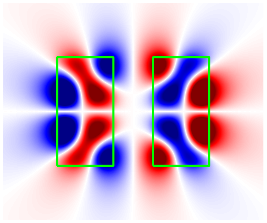} & $h_1/r=1.038, L_{12}/r=0.734$ & 2.19 & $5.7\cdot10^3$ & 1.9 \\
\hline
\newtag{7}&\includegraphics[height=1.2cm,clip=]{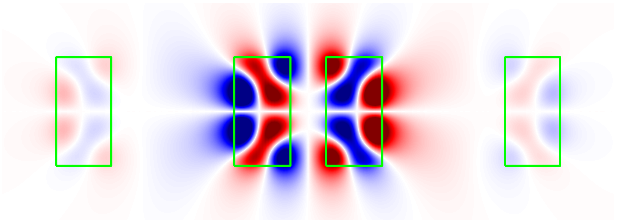} & $h_1/r=1.0173, r_1/r=1$\n$h_2/r=1.039, L_{12}/r=2.2731$\n$L_{23}/r=0.6585$ & 2.19 & $9.8\cdot10^5$ & 2 \\
\hline
\newtag{8}&\includegraphics[height=1.2cm,clip=]{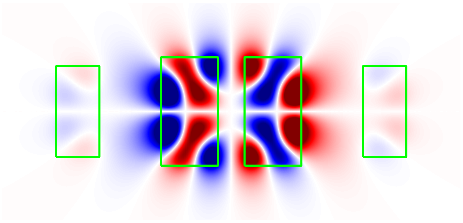} & $h_1/r=0.7988, r_1/r=0.8368$\n$h_2/r=1.0503, L_{12}/r=1.1424$\n$L_{23}/r=0.4922$ & 2.2 & $4.8\cdot10^6$ & 2.1 \\
\hline
\newtag{9}&\includegraphics[height=1.2cm,clip=]{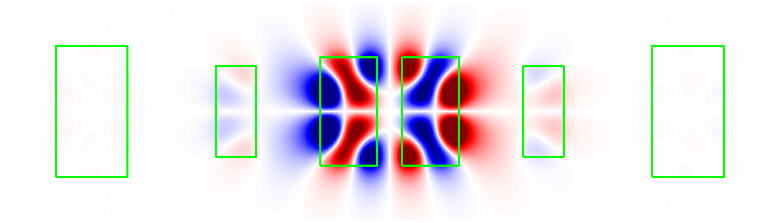} & $h_1/r=1.3257, r_1/r=1.2136$\n$h_2/r=0.7479, r_2/r=0.8376$\n$h_3/r=1.05326, L_{12}/r=1.6345$\n$L_{23}/r=1.1869, L_{34}/r=0.46103$ & 2.2 & $1.3\cdot10^9$ & 2.1 \\
\hline \hline
\newtag{10}&\includegraphics[height=2cm,clip=]{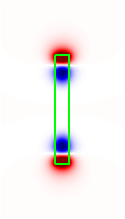}\includegraphics[height=2cm,clip=]{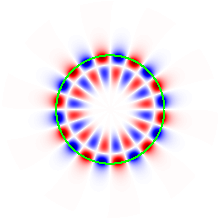} & $h_1/r=0.2588, m=10$ & 4.76 & $6\cdot10^6$ & 4.9 \\
\hline
\newtag{11}&\includegraphics[height=1.2cm,clip=]{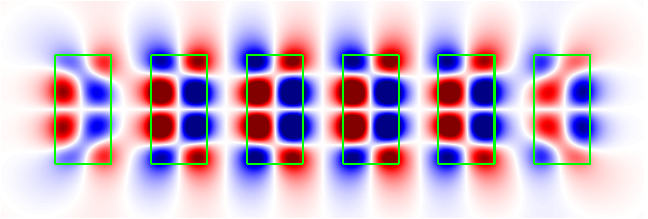} & $h_1=h_2=h_3=1.038r$\n$r_1=r_2=r$\n$L_{12}=L_{23}=L_{34}=0.734r$ & 2.19 & $2.4\cdot10^3$ & 4.4 \\
\hline
\end{tabular}
\caption{\label{Tab1}Mode profiles and parameters of optimized
systems of several coaxial disks.}
\end{table*}

These data in Table are compared with whispering gallery mode with
azimuthal number $m=10$ and the eigenfrequency ${\rm Re}(kr)=4.76$
in single disk of aspect ratio $h_1/r=0.2588$.

\section{Conclusion and Outlook}

The avoided crossing of resonances leads to substantial
redistribution of their imaginary parts and hybridization of
resonant modes \cite{Heiss2000}. This way of the $Q$-factor
enhancement was turned out to be successful even in a single
cavity shaped as disk \cite{Rybin2017} or long rod of rectangular
cross-section \cite{Huang2021}. The ACRs in two identical cavities
lifts the $Q$-factor essentially more
\cite{Wiersig2006,Boriskina2006,Bulgakov2021,Vennberg2022}. It
might be seemed, a further increasing of the number $N$ of
identical cavities is the best way to enhance $Q$-factor because
periodical array of cavities supports
quasi-BICs \cite{Sadrieva2019} with asymptotic $Q\sim N^2$.
However, this method of enhancement of $Q$-factor is bumping into
saturation owing to material losses \cite{Sadrieva2019} and
structural fluctuations \cite{Maslova2021}. Moreover, BIC modes
concede in compactness of DS and mode volume. In the present paper
we show that DS composed of cavities with different scales
provides considerably larger $Q$-factors preserving the mode
volume as Table \ref{Tab1} shows.
And what is remarkable, this unprecedented values of the
$Q$-factor refer to the compact DSs as crucially different from
the extended periodical DSs supporting the quasi-BICs. The
compactness of DS has a large technological advantage for sensing
and lasing devices.

There is a useful tool to understand the nature of the extremely
high quality factor for the avoided crossing through multipole
decomposition \cite{Jackson}. That tool shed light on the origin
of the high $Q$-factor in the isolated disk
\cite{Chen2019,Bogdanov2019} and the origin of bound states in the
continuum \cite{Sadrieva2019a}. In the present case of several
cavities we also observe that extreme $Q$-factor is attributed to
strong redistribution of radiation that originates from
compensation of dominating multipole coefficients. Moreover, we
show that it is related to almost perfect destructive interference
of the low order multipole radiations from the inner subsystem
inserted into the outer dimer.

Thus, 
the way to boost quality factor of the array of resonators looks
simple.
First, we attain the maximal $Q$-factor by ACRs in the inner
subsystem of $N-2$ disks that results in some hybridized resonant
mode $\psi_{s,a}(N-2)$, which can be symmetric or antisymmetric
relative to axis inversion. Then, we symmetrically enclose the
inner subsystem into a shell consisted of two identical disks,
which form an outer dimer.
Varying the scales of dimer (radius $r_1$, height $h_1$ and
distance $L_{12}$) we perform ACRs of resonant modes
$\psi_{s,a}(N-2)$ with the resonant modes $\psi_{s,a}(2)$ of the
outer dimer. To achieve an extremal $Q$-factors one has to allow
slight change of scales of inner subsystem too as it was shown in
the Section \ref{Sec3}.
Because of this, the optimization procedure should be performed
over all scales of the total system. As a result, we can achieve
almost perfect shielding of inner resonant mode by outer dimer and
boost the $Q$-factor by several orders of magnitude. Some
hybridized resonant modes
are collected in Table \ref{Tab1}. Our results show that
multiscale optimization procedure gives substantially higher
results for the $Q$-factor compared to the case of equidistant
identical disks, which supports quasi-BICs at $\Gamma$-point.

The proposed algorithm could be easily adopted to multi cavity
systems of different shape and permittivity. The present system of
coaxial disks was chosen because of separation of polarizations in
the sector of zero azimuthal index. It should be noted that the
$Q$-factor of optimized systems is very sensitive to scale
parameters however. Especially to parameters, which affect the
resonant mode $\psi_{N-2}$ of the inner subsystem because of weak
localization of the total mode $\psi_N$ at outside dimer. This
brings definite technological problems because of necessity to
accurately set up different scales of different resonators.
Fortunately, S. Kim {\it et al} \cite{Kim2019} reported control
over disk dimensions with accuracy 5 nm in optical range, i.e.,
0.3 percent relative to the optical wave length $\lambda=1.55 \mu
m$.

{\bf Acknowledgement}: We acknowledge discussions with Yi Xu.

{\bf Funding}: This work is supported by the Russian Science
Foundation under grant 22-12-00070.

{\bf Conflict of interest statement}: The authors declare no
conflicts of interest regarding this article.

\begin{thebibliography}{66}%
\makeatletter
\providecommand \@ifxundefined [1]{%
 \@ifx{#1\undefined}
}%
\providecommand \@ifnum [1]{%
 \ifnum #1\expandafter \@firstoftwo
 \else \expandafter \@secondoftwo
 \fi
}%
\providecommand \@ifx [1]{%
 \ifx #1\expandafter \@firstoftwo
 \else \expandafter \@secondoftwo
 \fi
}%
\providecommand \natexlab [1]{#1}%
\providecommand \enquote  [1]{``#1''}%
\providecommand \bibnamefont  [1]{#1}%
\providecommand \bibfnamefont [1]{#1}%
\providecommand \citenamefont [1]{#1}%
\providecommand \href@noop [0]{\@secondoftwo}%
\providecommand \href [0]{\begingroup \@sanitize@url \@href}%
\providecommand \@href[1]{\@@startlink{#1}\@@href}%
\providecommand \@@href[1]{\endgroup#1\@@endlink}%
\providecommand \@sanitize@url [0]{\catcode `\\12\catcode `\$12\catcode
  `\&12\catcode `\#12\catcode `\^12\catcode `\_12\catcode `\%12\relax}%
\providecommand \@@startlink[1]{}%
\providecommand \@@endlink[0]{}%
\providecommand \url  [0]{\begingroup\@sanitize@url \@url }%
\providecommand \@url [1]{\endgroup\@href {#1}{\urlprefix }}%
\providecommand \urlprefix  [0]{URL }%
\providecommand \Eprint [0]{\href }%
\providecommand \doibase [0]{https://doi.org/}%
\providecommand \selectlanguage [0]{\@gobble}%
\providecommand \bibinfo  [0]{\@secondoftwo}%
\providecommand \bibfield  [0]{\@secondoftwo}%
\providecommand \translation [1]{[#1]}%
\providecommand \BibitemOpen [0]{}%
\providecommand \bibitemStop [0]{}%
\providecommand \bibitemNoStop [0]{.\EOS\space}%
\providecommand \EOS [0]{\spacefactor3000\relax}%
\providecommand \BibitemShut  [1]{\csname bibitem#1\endcsname}%
\let\auto@bib@innerbib\@empty
\bibitem [{\citenamefont {Mie}(1908)}]{Mie1908}%
  \BibitemOpen
  \bibfield  {author} {\bibinfo {author} {\bibfnamefont {G.}~\bibnamefont
  {Mie}},\ }\bibfield  {title} {\enquote {\bibinfo {title} {Beitr{\"a}ge zur
  optik tr{\"u}ber medien, speziell kolloidaler metall{\"o}sungen},}\ }\href
  {https://doi.org/10.1002/andp.19083300302} {\bibfield  {journal} {\bibinfo
  {journal} {Annalen der Physik}\ }\textbf {\bibinfo {volume} {330}},\ \bibinfo
  {pages} {377--445} (\bibinfo {year} {1908})}\BibitemShut {NoStop}%
\bibitem [{\citenamefont {Colton}\ and\ \citenamefont
  {R.Kress}(1998)}]{Colton}%
  \BibitemOpen
  \bibfield  {author} {\bibinfo {author} {\bibfnamefont {D.}~\bibnamefont
  {Colton}}\ and\ \bibinfo {author} {\bibnamefont {R.Kress}},\ }\href@noop {}
  {\emph {\bibinfo {title} {Inverse Acoustic and Electromagnetic Scattering
  Theory}}},\ \bibinfo {edition} {2nd}\ ed.\ (\bibinfo  {publisher}
  {Springer,Berlin},\ \bibinfo {year} {1998})\BibitemShut {NoStop}%
\bibitem [{\citenamefont {Silveirinha}(2014)}]{Silveirinha14}%
  \BibitemOpen
  \bibfield  {author} {\bibinfo {author} {\bibfnamefont {M.~G.}\ \bibnamefont
  {Silveirinha}},\ }\bibfield  {title} {\enquote {\bibinfo {title} {Trapping
  light in open plasmonic nanostructures},}\ }\href@noop {} {\bibfield
  {journal} {\bibinfo  {journal} {Phys. Rev. A}\ }\textbf {\bibinfo {volume}
  {89}},\ \bibinfo {pages} {023813} (\bibinfo {year} {2014})}\BibitemShut
  {NoStop}%
\bibitem [{\citenamefont {Vahala}(2003)}]{Vahala2003}%
  \BibitemOpen
  \bibfield  {author} {\bibinfo {author} {\bibfnamefont {K.}~\bibnamefont
  {Vahala}},\ }\bibfield  {title} {\enquote {\bibinfo {title} {Optical
  microcavities},}\ }\href {https://doi.org/10.1038/nature01939} {\bibfield
  {journal} {\bibinfo  {journal} {Nature}\ }\textbf {\bibinfo {volume} {424}},\
  \bibinfo {pages} {839--846} (\bibinfo {year} {2003})}\BibitemShut {NoStop}%
\bibitem [{\citenamefont {Ryckman}\ and\ \citenamefont
  {Weiss}(2012)}]{Ryckman2012}%
  \BibitemOpen
  \bibfield  {author} {\bibinfo {author} {\bibfnamefont {J.}~\bibnamefont
  {Ryckman}}\ and\ \bibinfo {author} {\bibfnamefont {S.~M.}\ \bibnamefont
  {Weiss}},\ }\bibfield  {title} {\enquote {\bibinfo {title} {Low mode volume
  slotted photonic crystal single nanobeam cavity},}\ }\href
  {https://doi.org/10.1063/1.4742749} {\bibfield  {journal} {\bibinfo
  {journal} {Appl. Phys. Lett.}\ }\textbf {\bibinfo {volume} {101}},\ \bibinfo
  {pages} {071104} (\bibinfo {year} {2012})}\BibitemShut {NoStop}%
\bibitem [{\citenamefont {Seidler}\ \emph {et~al.}(2013)\citenamefont
  {Seidler}, \citenamefont {Lister}, \citenamefont {Drechsler}, \citenamefont
  {Hofrichter},\ and\ \citenamefont {St{\"o}ferle}}]{Seidler2013}%
  \BibitemOpen
  \bibfield  {author} {\bibinfo {author} {\bibfnamefont {P.}~\bibnamefont
  {Seidler}}, \bibinfo {author} {\bibfnamefont {K.}~\bibnamefont {Lister}},
  \bibinfo {author} {\bibfnamefont {U.}~\bibnamefont {Drechsler}}, \bibinfo
  {author} {\bibfnamefont {J.}~\bibnamefont {Hofrichter}},\ and\ \bibinfo
  {author} {\bibfnamefont {T.}~\bibnamefont {St{\"o}ferle}},\ }\bibfield
  {title} {\enquote {\bibinfo {title} {Slotted photonic crystal nanobeam cavity
  with an ultrahigh quality factor-to-mode volume ratio},}\ }\href
  {https://doi.org/10.1364/oe.21.032468} {\bibfield  {journal} {\bibinfo
  {journal} {Optics Express}\ }\textbf {\bibinfo {volume} {21}},\ \bibinfo
  {pages} {32468} (\bibinfo {year} {2013})}\BibitemShut {NoStop}%
\bibitem [{\citenamefont {Zhou}\ \emph {et~al.}(2019)\citenamefont {Zhou},
  \citenamefont {Zheng}, \citenamefont {Fang}, \citenamefont {Xu},\ and\
  \citenamefont {Majumdar}}]{Zhou2019}%
  \BibitemOpen
  \bibfield  {author} {\bibinfo {author} {\bibfnamefont {J.}~\bibnamefont
  {Zhou}}, \bibinfo {author} {\bibfnamefont {J.}~\bibnamefont {Zheng}},
  \bibinfo {author} {\bibfnamefont {Z.}~\bibnamefont {Fang}}, \bibinfo {author}
  {\bibfnamefont {P.}~\bibnamefont {Xu}},\ and\ \bibinfo {author}
  {\bibfnamefont {A.}~\bibnamefont {Majumdar}},\ }\bibfield  {title} {\enquote
  {\bibinfo {title} {Ultra-low mode volume on substrate silicon nanobeam
  cavity},}\ }\href {https://doi.org/10.1364/oe.27.030692} {\bibfield
  {journal} {\bibinfo  {journal} {Optics Express}\ }\textbf {\bibinfo {volume}
  {27}},\ \bibinfo {pages} {30692} (\bibinfo {year} {2019})}\BibitemShut
  {NoStop}%
\bibitem [{\citenamefont {Braginsky}, \citenamefont {Gorodetsky},\ and\
  \citenamefont {Ilchenko}(1989)}]{Braginsky1989}%
  \BibitemOpen
  \bibfield  {author} {\bibinfo {author} {\bibfnamefont {V.}~\bibnamefont
  {Braginsky}}, \bibinfo {author} {\bibfnamefont {M.}~\bibnamefont
  {Gorodetsky}},\ and\ \bibinfo {author} {\bibfnamefont {V.}~\bibnamefont
  {Ilchenko}},\ }\bibfield  {title} {\enquote {\bibinfo {title} {Quality-factor
  and nonlinear properties of optical whispering-gallery modes},}\ }\href
  {https://doi.org/10.1016/0375-9601(89)90912-2} {\bibfield  {journal}
  {\bibinfo  {journal} {Phys. Let. A}\ }\textbf {\bibinfo {volume} {137}},\
  \bibinfo {pages} {393} (\bibinfo {year} {1989})}\BibitemShut {NoStop}%
\bibitem [{\citenamefont {Gorodetsky}\ and\ \citenamefont
  {Fomin}(2006)}]{Gorodetsky2006}%
  \BibitemOpen
  \bibfield  {author} {\bibinfo {author} {\bibfnamefont {M.}~\bibnamefont
  {Gorodetsky}}\ and\ \bibinfo {author} {\bibfnamefont {A.}~\bibnamefont
  {Fomin}},\ }\bibfield  {title} {\enquote {\bibinfo {title} {Geometrical
  theory of whispering-gallery modes},}\ }\href
  {https://doi.org/10.1109/jstqe.2005.862954} {\bibfield  {journal} {\bibinfo
  {journal} {{IEEE} J. Selected Topics in Quantum Electronics}\ }\textbf
  {\bibinfo {volume} {12}},\ \bibinfo {pages} {33--39} (\bibinfo {year}
  {2006})}\BibitemShut {NoStop}%
\bibitem [{\citenamefont {Acharyya}\ and\ \citenamefont
  {Kozyreff}(2019)}]{Acharyya2019}%
  \BibitemOpen
  \bibfield  {author} {\bibinfo {author} {\bibfnamefont {N.}~\bibnamefont
  {Acharyya}}\ and\ \bibinfo {author} {\bibfnamefont {G.}~\bibnamefont
  {Kozyreff}},\ }\bibfield  {title} {\enquote {\bibinfo {title} {Large q factor
  with very small whispering gallery mode resonators},}\ }\href
  {https://doi.org/10.1103/physrevapplied.12.014060} {\bibfield  {journal}
  {\bibinfo  {journal} {Phys. Rev. Appl.}\ }\textbf {\bibinfo {volume} {12}},\
  \bibinfo {pages} {014060} (\bibinfo {year} {2019})}\BibitemShut {NoStop}%
\bibitem [{\citenamefont {Jiang}\ \emph {et~al.}(2020)\citenamefont {Jiang},
  \citenamefont {Qavi}, \citenamefont {Huang},\ and\ \citenamefont
  {Yang}}]{Jiang2020}%
  \BibitemOpen
  \bibfield  {author} {\bibinfo {author} {\bibfnamefont {X.}~\bibnamefont
  {Jiang}}, \bibinfo {author} {\bibfnamefont {A.}~\bibnamefont {Qavi}},
  \bibinfo {author} {\bibfnamefont {S.}~\bibnamefont {Huang}},\ and\ \bibinfo
  {author} {\bibfnamefont {L.}~\bibnamefont {Yang}},\ }\bibfield  {title}
  {\enquote {\bibinfo {title} {Whispering-gallery sensors},}\ }\href
  {https://doi.org/10.1016/j.matt.2020.07.008} {\bibfield  {journal} {\bibinfo
  {journal} {Matter}\ }\textbf {\bibinfo {volume} {3}},\ \bibinfo {pages}
  {371--392} (\bibinfo {year} {2020})}\BibitemShut {NoStop}%
\bibitem [{\citenamefont {Hong}\ \emph {et~al.}(2022)\citenamefont {Hong},
  \citenamefont {Lee}, \citenamefont {Hong}, \citenamefont {Kim},\ and\
  \citenamefont {Kwon}}]{Hong2022}%
  \BibitemOpen
  \bibfield  {author} {\bibinfo {author} {\bibfnamefont {S.~H.}\ \bibnamefont
  {Hong}}, \bibinfo {author} {\bibfnamefont {Y.~J.}\ \bibnamefont {Lee}},
  \bibinfo {author} {\bibfnamefont {S.}~\bibnamefont {Hong}}, \bibinfo {author}
  {\bibfnamefont {Y.}~\bibnamefont {Kim}},\ and\ \bibinfo {author}
  {\bibfnamefont {S.-H.}\ \bibnamefont {Kwon}},\ }\bibfield  {title} {\enquote
  {\bibinfo {title} {Antisymmetric mode cancellation for high-{Q} cavities in a
  double disk},}\ }\href {https://doi.org/10.3390/photonics9080572} {\bibfield
  {journal} {\bibinfo  {journal} {Photonics}\ }\textbf {\bibinfo {volume}
  {9}},\ \bibinfo {pages} {572} (\bibinfo {year} {2022})}\BibitemShut {NoStop}%
\bibitem [{\citenamefont {{Chia Wei Hsu}}\ \emph {et~al.}(2016)\citenamefont
  {{Chia Wei Hsu}}, \citenamefont {{Bo Zhen}}, \citenamefont {Stone},
  \citenamefont {Joannopoulos},\ and\ \citenamefont
  {Solja{\v{c}}i{\'{c}}}}]{Hsu2016}%
  \BibitemOpen
  \bibfield  {author} {\bibinfo {author} {\bibnamefont {{Chia Wei Hsu}}},
  \bibinfo {author} {\bibnamefont {{Bo Zhen}}}, \bibinfo {author}
  {\bibfnamefont {A.~D.}\ \bibnamefont {Stone}}, \bibinfo {author}
  {\bibfnamefont {J.~D.}\ \bibnamefont {Joannopoulos}},\ and\ \bibinfo {author}
  {\bibfnamefont {M.}~\bibnamefont {Solja{\v{c}}i{\'{c}}}},\ }\bibfield
  {title} {\enquote {\bibinfo {title} {Bound states in the continuum},}\ }\href
  {https://doi.org/10.1038/natrevmats.2016.48} {\bibfield  {journal} {\bibinfo
  {journal} {Nature Reviews Materials}\ }\textbf {\bibinfo {volume} {1}},\
  \bibinfo {pages} {16048} (\bibinfo {year} {2016})}\BibitemShut {NoStop}%
\bibitem [{\citenamefont {Azzam}\ and\ \citenamefont
  {Kildishev}(2020)}]{Azzam2020}%
  \BibitemOpen
  \bibfield  {author} {\bibinfo {author} {\bibfnamefont {S.}~\bibnamefont
  {Azzam}}\ and\ \bibinfo {author} {\bibfnamefont {A.}~\bibnamefont
  {Kildishev}},\ }\bibfield  {title} {\enquote {\bibinfo {title} {Photonic
  bound states in the continuum: From basics to applications},}\ }\href
  {https://doi.org/10.1002/adom.202001469} {\bibfield  {journal} {\bibinfo
  {journal} {Adv. Opt. Mat.}\ }\textbf {\bibinfo {volume} {9}},\ \bibinfo
  {pages} {2001469} (\bibinfo {year} {2020})}\BibitemShut {NoStop}%
\bibitem [{\citenamefont {Huang}\ \emph {et~al.}(2020)\citenamefont {Huang},
  \citenamefont {Xu}, \citenamefont {Woolley},\ and\ \citenamefont
  {Miroshnichenko}}]{Huang2020}%
  \BibitemOpen
  \bibfield  {author} {\bibinfo {author} {\bibfnamefont {L.}~\bibnamefont
  {Huang}}, \bibinfo {author} {\bibfnamefont {L.}~\bibnamefont {Xu}}, \bibinfo
  {author} {\bibfnamefont {M.}~\bibnamefont {Woolley}},\ and\ \bibinfo {author}
  {\bibfnamefont {A.~E.}\ \bibnamefont {Miroshnichenko}},\ }\bibfield  {title}
  {\enquote {\bibinfo {title} {Trends in quantum nanophotonics},}\ }\href
  {https://doi.org/10.1002/qute.201900126} {\bibfield  {journal} {\bibinfo
  {journal} {Advanced Quantum Technologies}\ }\textbf {\bibinfo {volume} {3}},\
  \bibinfo {pages} {1900126} (\bibinfo {year} {2020})}\BibitemShut {NoStop}%
\bibitem [{\citenamefont {Joseph}\ \emph {et~al.}(2021)\citenamefont {Joseph},
  \citenamefont {Pandey}, \citenamefont {Sarkar},\ and\ \citenamefont
  {Joseph}}]{Joseph2021}%
  \BibitemOpen
  \bibfield  {author} {\bibinfo {author} {\bibfnamefont {S.}~\bibnamefont
  {Joseph}}, \bibinfo {author} {\bibfnamefont {S.}~\bibnamefont {Pandey}},
  \bibinfo {author} {\bibfnamefont {S.}~\bibnamefont {Sarkar}},\ and\ \bibinfo
  {author} {\bibfnamefont {J.}~\bibnamefont {Joseph}},\ }\bibfield  {title}
  {\enquote {\bibinfo {title} {Bound states in the continuum in resonant
  nanostructures: an overview of engineered materials for tailored
  applications},}\ }\href {https://doi.org/10.1515/nanoph-2021-0387} {\bibfield
   {journal} {\bibinfo  {journal} {Nanophotonics}\ }\textbf {\bibinfo {volume}
  {10}},\ \bibinfo {pages} {4175--4207} (\bibinfo {year} {2021})}\BibitemShut
  {NoStop}%
\bibitem [{\citenamefont {Koshelev}\ \emph {et~al.}(2022)\citenamefont
  {Koshelev}, \citenamefont {Sadrieva}, \citenamefont {Shcherbakov},
  \citenamefont {Kivshar},\ and\ \citenamefont {Bogdanov}}]{Koshelev2022}%
  \BibitemOpen
  \bibfield  {author} {\bibinfo {author} {\bibfnamefont {K.}~\bibnamefont
  {Koshelev}}, \bibinfo {author} {\bibfnamefont {Z.}~\bibnamefont {Sadrieva}},
  \bibinfo {author} {\bibfnamefont {A.}~\bibnamefont {Shcherbakov}}, \bibinfo
  {author} {\bibfnamefont {Y.}~\bibnamefont {Kivshar}},\ and\ \bibinfo {author}
  {\bibfnamefont {A.}~\bibnamefont {Bogdanov}},\ }\bibfield  {title} {\enquote
  {\bibinfo {title} {Bound states in the continuum in photonic structures},}\
  }\href {https://doi.org/10.3367/UFNe.2021.12.039120} {\bibfield  {journal}
  {\bibinfo  {journal} {Uspekhi Fizicheskikh Nauk}\ }\textbf {\bibinfo {volume}
  {65}},\ \bibinfo {pages} {039120} (\bibinfo {year} {2022})},\ \Eprint
  {https://arxiv.org/abs/2207.01441v2} {2207.01441v2} \BibitemShut {NoStop}%
\bibitem [{\citenamefont {Hu}\ \emph {et~al.}(2022)\citenamefont {Hu},
  \citenamefont {Xie}, \citenamefont {Song}, \citenamefont {Chen},
  \citenamefont {Xiang}, \citenamefont {Han},\ and\ \citenamefont
  {Zi}}]{Hu2022}%
  \BibitemOpen
  \bibfield  {author} {\bibinfo {author} {\bibfnamefont {P.}~\bibnamefont
  {Hu}}, \bibinfo {author} {\bibfnamefont {C.}~\bibnamefont {Xie}}, \bibinfo
  {author} {\bibfnamefont {Q.}~\bibnamefont {Song}}, \bibinfo {author}
  {\bibfnamefont {A.}~\bibnamefont {Chen}}, \bibinfo {author} {\bibfnamefont
  {H.}~\bibnamefont {Xiang}}, \bibinfo {author} {\bibfnamefont
  {D.}~\bibnamefont {Han}},\ and\ \bibinfo {author} {\bibfnamefont
  {J.}~\bibnamefont {Zi}},\ }\bibfield  {title} {\enquote {\bibinfo {title}
  {Bound states in the continuum based on the total internal reflection of
  bloch waves},}\ }\href {https://doi.org/10.1093/nsr/nwac043} {\bibfield
  {journal} {\bibinfo  {journal} {Nat. Sci. Review}\ } (\bibinfo {year}
  {2022}),\ 10.1093/nsr/nwac043}\BibitemShut {NoStop}%
\bibitem [{\citenamefont {{Chia Wei Hsu}}\ \emph {et~al.}(2013)\citenamefont
  {{Chia Wei Hsu}}, \citenamefont {{Bo Zhen}}, \citenamefont {{Jeongwon Lee}},
  \citenamefont {Johnson}, \citenamefont {Joannopoulos},\ and\ \citenamefont
  {Solja{\v{c}}i{\'c}}}]{Hsu2013}%
  \BibitemOpen
  \bibfield  {author} {\bibinfo {author} {\bibnamefont {{Chia Wei Hsu}}},
  \bibinfo {author} {\bibnamefont {{Bo Zhen}}}, \bibinfo {author} {\bibnamefont
  {{Jeongwon Lee}}}, \bibinfo {author} {\bibfnamefont {S.~G.}\ \bibnamefont
  {Johnson}}, \bibinfo {author} {\bibfnamefont {J.~D.}\ \bibnamefont
  {Joannopoulos}},\ and\ \bibinfo {author} {\bibfnamefont {M.}~\bibnamefont
  {Solja{\v{c}}i{\'c}}},\ }\bibfield  {title} {\enquote {\bibinfo {title}
  {Observation of trapped light within the radiation continuum},}\ }\href
  {https://doi.org/10.1038/nature12289} {\bibfield  {journal} {\bibinfo
  {journal} {Nature}\ }\textbf {\bibinfo {volume} {499}},\ \bibinfo {pages}
  {188} (\bibinfo {year} {2013})}\BibitemShut {NoStop}%
\bibitem [{\citenamefont {Bulgakov}\ and\ \citenamefont
  {Sadreev}(2017{\natexlab{a}})}]{Bulgakov2017}%
  \BibitemOpen
  \bibfield  {author} {\bibinfo {author} {\bibfnamefont {E.}~\bibnamefont
  {Bulgakov}}\ and\ \bibinfo {author} {\bibfnamefont {A.}~\bibnamefont
  {Sadreev}},\ }\bibfield  {title} {\enquote {\bibinfo {title} {Bound states in
  the continuum with high orbital angular momentum in a dielectric rod with
  periodically modulated permittivity},}\ }\href
  {https://doi.org/10.1103/physreva.96.013841} {\bibfield  {journal} {\bibinfo
  {journal} {Phys. Rev. A}\ }\textbf {\bibinfo {volume} {96}},\ \bibinfo
  {pages} {013841} (\bibinfo {year} {2017}{\natexlab{a}})}\BibitemShut
  {NoStop}%
\bibitem [{\citenamefont {Koshelev}\ \emph {et~al.}(2019)\citenamefont
  {Koshelev}, \citenamefont {Favraud}, \citenamefont {Bogdanov}, \citenamefont
  {Kivshar},\ and\ \citenamefont {Fratalocchi}}]{Koshelev2019}%
  \BibitemOpen
  \bibfield  {author} {\bibinfo {author} {\bibfnamefont {K.}~\bibnamefont
  {Koshelev}}, \bibinfo {author} {\bibfnamefont {G.}~\bibnamefont {Favraud}},
  \bibinfo {author} {\bibfnamefont {A.}~\bibnamefont {Bogdanov}}, \bibinfo
  {author} {\bibfnamefont {Y.}~\bibnamefont {Kivshar}},\ and\ \bibinfo {author}
  {\bibfnamefont {A.}~\bibnamefont {Fratalocchi}},\ }\bibfield  {title}
  {\enquote {\bibinfo {title} {Nonradiating photonics with resonant dielectric
  nanostructures},}\ }\href {https://doi.org/10.1515/nanoph-2019-0024}
  {\bibfield  {journal} {\bibinfo  {journal} {Nanophotonics}\ }\textbf
  {\bibinfo {volume} {8}},\ \bibinfo {pages} {725} (\bibinfo {year}
  {2019})}\BibitemShut {NoStop}%
\bibitem [{\citenamefont {Taghizadeh}\ and\ \citenamefont
  {Chung}(2017)}]{Taghizadeh2017}%
  \BibitemOpen
  \bibfield  {author} {\bibinfo {author} {\bibfnamefont {A.}~\bibnamefont
  {Taghizadeh}}\ and\ \bibinfo {author} {\bibfnamefont {I.-S.}\ \bibnamefont
  {Chung}},\ }\bibfield  {title} {\enquote {\bibinfo {title} {Quasi bound
  states in the continuum with few unit cells of photonic crystal slab},}\
  }\href {https://doi.org/10.1063/1.4990753} {\bibfield  {journal} {\bibinfo
  {journal} {Appl. Phys. Lett.}\ }\textbf {\bibinfo {volume} {111}},\ \bibinfo
  {pages} {031114} (\bibinfo {year} {2017})}\BibitemShut {NoStop}%
\bibitem [{\citenamefont {Bulgakov}\ and\ \citenamefont
  {Sadreev}(2017{\natexlab{b}})}]{Bulgakov2017b}%
  \BibitemOpen
  \bibfield  {author} {\bibinfo {author} {\bibfnamefont {E.}~\bibnamefont
  {Bulgakov}}\ and\ \bibinfo {author} {\bibfnamefont {A.}~\bibnamefont
  {Sadreev}},\ }\bibfield  {title} {\enquote {\bibinfo {title} {Propagating
  bloch bound states with orbital angular momentum above the light line in the
  array of dielectric spheres},}\ }\href
  {https://doi.org/10.1364/josaa.34.000949} {\bibfield  {journal} {\bibinfo
  {journal} {J. Opt. Soc. Am. A}\ }\textbf {\bibinfo {volume} {34}},\ \bibinfo
  {pages} {949} (\bibinfo {year} {2017}{\natexlab{b}})}\BibitemShut {NoStop}%
\bibitem [{\citenamefont {Sadrieva}\ \emph
  {et~al.}(2019{\natexlab{a}})\citenamefont {Sadrieva}, \citenamefont
  {Belyakov}, \citenamefont {Balezin}, \citenamefont {Kapitanova},
  \citenamefont {Nenasheva}, \citenamefont {Sadreev},\ and\ \citenamefont
  {Bogdanov}}]{Sadrieva2019}%
  \BibitemOpen
  \bibfield  {author} {\bibinfo {author} {\bibfnamefont {Z.~F.}\ \bibnamefont
  {Sadrieva}}, \bibinfo {author} {\bibfnamefont {M.~A.}\ \bibnamefont
  {Belyakov}}, \bibinfo {author} {\bibfnamefont {M.~A.}\ \bibnamefont
  {Balezin}}, \bibinfo {author} {\bibfnamefont {P.~V.}\ \bibnamefont
  {Kapitanova}}, \bibinfo {author} {\bibfnamefont {E.~A.}\ \bibnamefont
  {Nenasheva}}, \bibinfo {author} {\bibfnamefont {A.~F.}\ \bibnamefont
  {Sadreev}},\ and\ \bibinfo {author} {\bibfnamefont {A.~A.}\ \bibnamefont
  {Bogdanov}},\ }\bibfield  {title} {\enquote {\bibinfo {title} {Experimental
  observation of a symmetry-protected bound state in the continuum in a chain
  of dielectric disks},}\ }\href {https://doi.org/10.1103/physreva.99.053804}
  {\bibfield  {journal} {\bibinfo  {journal} {Phys. Rev. A}\ }\textbf {\bibinfo
  {volume} {99}},\ \bibinfo {pages} {053804} (\bibinfo {year}
  {2019}{\natexlab{a}})}\BibitemShut {NoStop}%
\bibitem [{\citenamefont {Polishchuk}\ \emph {et~al.}(2017)\citenamefont
  {Polishchuk}, \citenamefont {Anastasiev}, \citenamefont {Tsyvkunova},
  \citenamefont {Gozman}, \citenamefont {Solov'ov},\ and\ \citenamefont
  {Polishchuk}}]{Polishchuk}%
  \BibitemOpen
  \bibfield  {author} {\bibinfo {author} {\bibfnamefont {I.~Y.}\ \bibnamefont
  {Polishchuk}}, \bibinfo {author} {\bibfnamefont {A.~A.}\ \bibnamefont
  {Anastasiev}}, \bibinfo {author} {\bibfnamefont {E.~A.}\ \bibnamefont
  {Tsyvkunova}}, \bibinfo {author} {\bibfnamefont {M.~I.}\ \bibnamefont
  {Gozman}}, \bibinfo {author} {\bibfnamefont {S.~V.}\ \bibnamefont
  {Solov'ov}},\ and\ \bibinfo {author} {\bibfnamefont {Y.~I.}\ \bibnamefont
  {Polishchuk}},\ }\bibfield  {title} {\enquote {\bibinfo {title} {Guided modes
  in the plane array of optical waveguides},}\ }\href
  {https://doi.org/10.1103/physreva.95.053847} {\bibfield  {journal} {\bibinfo
  {journal} {Phys. Rev. A}\ }\textbf {\bibinfo {volume} {95}},\ \bibinfo
  {pages} {053847} (\bibinfo {year} {2017})}\BibitemShut {NoStop}%
\bibitem [{\citenamefont {Sidorenko}\ \emph {et~al.}(2021)\citenamefont
  {Sidorenko}, \citenamefont {Sergaeva}, \citenamefont {Sadrieva},
  \citenamefont {Roques-Carmes}, \citenamefont {Muraev}, \citenamefont
  {Maksimov},\ and\ \citenamefont {Bogdanov}}]{Sidorenko2021}%
  \BibitemOpen
  \bibfield  {author} {\bibinfo {author} {\bibfnamefont {M.}~\bibnamefont
  {Sidorenko}}, \bibinfo {author} {\bibfnamefont {O.}~\bibnamefont {Sergaeva}},
  \bibinfo {author} {\bibfnamefont {Z.}~\bibnamefont {Sadrieva}}, \bibinfo
  {author} {\bibfnamefont {C.}~\bibnamefont {Roques-Carmes}}, \bibinfo {author}
  {\bibfnamefont {P.}~\bibnamefont {Muraev}}, \bibinfo {author} {\bibfnamefont
  {D.}~\bibnamefont {Maksimov}},\ and\ \bibinfo {author} {\bibfnamefont
  {A.}~\bibnamefont {Bogdanov}},\ }\bibfield  {title} {\enquote {\bibinfo
  {title} {Observation of an accidental bound state in the continuum in a chain
  of dielectric disks},}\ }\href
  {https://doi.org/10.1103/physrevapplied.15.034041} {\bibfield  {journal}
  {\bibinfo  {journal} {Phys. Rev. Appl.}\ }\textbf {\bibinfo {volume} {15}},\
  \bibinfo {pages} {034041} (\bibinfo {year} {2021})}\BibitemShut {NoStop}%
\bibitem [{\citenamefont {Bulgakov}\ and\ \citenamefont
  {Sadreev}(2019)}]{Bulgakov2019a}%
  \BibitemOpen
  \bibfield  {author} {\bibinfo {author} {\bibfnamefont {E.}~\bibnamefont
  {Bulgakov}}\ and\ \bibinfo {author} {\bibfnamefont {A.}~\bibnamefont
  {Sadreev}},\ }\bibfield  {title} {\enquote {\bibinfo {title} {High-{Q}
  resonant modes in a finite array of dielectric particles},}\ }\href
  {https://doi.org/10.1103/physreva.99.033851} {\bibfield  {journal} {\bibinfo
  {journal} {Phys. Rev. A}\ }\textbf {\bibinfo {volume} {99}},\ \bibinfo
  {pages} {033851} (\bibinfo {year} {2019})}\BibitemShut {NoStop}%
\bibitem [{\citenamefont {Bulgakov}\ and\ \citenamefont
  {Maksimov}(2019)}]{Bulgakov2019b}%
  \BibitemOpen
  \bibfield  {author} {\bibinfo {author} {\bibfnamefont {E.}~\bibnamefont
  {Bulgakov}}\ and\ \bibinfo {author} {\bibfnamefont {D.}~\bibnamefont
  {Maksimov}},\ }\bibfield  {title} {\enquote {\bibinfo {title} {Q -factor
  optimization in dielectric oligomers},}\ }\href
  {https://doi.org/10.1103/physreva.100.033830} {\bibfield  {journal} {\bibinfo
   {journal} {Phys. Rev. A}\ }\textbf {\bibinfo {volume} {100}},\ \bibinfo
  {pages} {033830} (\bibinfo {year} {2019})}\BibitemShut {NoStop}%
\bibitem [{\citenamefont {Kornovan}\ \emph {et~al.}(2021)\citenamefont
  {Kornovan}, \citenamefont {Savelev}, \citenamefont {Kivshar},\ and\
  \citenamefont {Petrov}}]{Kornovan2021}%
  \BibitemOpen
  \bibfield  {author} {\bibinfo {author} {\bibfnamefont {D.~F.}\ \bibnamefont
  {Kornovan}}, \bibinfo {author} {\bibfnamefont {R.~S.}\ \bibnamefont
  {Savelev}}, \bibinfo {author} {\bibfnamefont {Y.}~\bibnamefont {Kivshar}},\
  and\ \bibinfo {author} {\bibfnamefont {M.~I.}\ \bibnamefont {Petrov}},\
  }\bibfield  {title} {\enquote {\bibinfo {title} {High-{Q} localized states in
  finite arrays of subwavelength resonators},}\ }\href
  {https://doi.org/10.1021/acsphotonics.1c01262} {\bibfield  {journal}
  {\bibinfo  {journal} {{ACS} Photonics}\ }\textbf {\bibinfo {volume} {8}},\
  \bibinfo {pages} {3627--3632} (\bibinfo {year} {2021})}\BibitemShut {NoStop}%
\bibitem [{\citenamefont {Gao}\ and\ \citenamefont {Shi}(2019)}]{Gao2019}%
  \BibitemOpen
  \bibfield  {author} {\bibinfo {author} {\bibfnamefont {Y.}~\bibnamefont
  {Gao}}\ and\ \bibinfo {author} {\bibfnamefont {Y.}~\bibnamefont {Shi}},\
  }\bibfield  {title} {\enquote {\bibinfo {title} {Design of a single
  nanoparticle trapping device based on bow-tie-shaped photonic crystal
  nanobeam cavities},}\ }\href {https://doi.org/10.1109/jphot.2019.2911291}
  {\bibfield  {journal} {\bibinfo  {journal} {{IEEE} Photonics Journal}\
  }\textbf {\bibinfo {volume} {11}},\ \bibinfo {pages} {1--8} (\bibinfo {year}
  {2019})}\BibitemShut {NoStop}%
\bibitem [{\citenamefont {von Brentano}(1990)}]{Brentano1990}%
  \BibitemOpen
  \bibfield  {author} {\bibinfo {author} {\bibfnamefont {P.}~\bibnamefont {von
  Brentano}},\ }\bibfield  {title} {\enquote {\bibinfo {title} {On the
  generalization of the level repulsion theorem to resonances},}\ }\href
  {https://doi.org/10.1016/0370-2693(90)92089-2} {\bibfield  {journal}
  {\bibinfo  {journal} {Phys. Lett. B}\ }\textbf {\bibinfo {volume} {238}},\
  \bibinfo {pages} {1} (\bibinfo {year} {1990})}\BibitemShut {NoStop}%
\bibitem [{\citenamefont {Heiss}(2000)}]{Heiss2000}%
  \BibitemOpen
  \bibfield  {author} {\bibinfo {author} {\bibfnamefont {W.~D.}\ \bibnamefont
  {Heiss}},\ }\bibfield  {title} {\enquote {\bibinfo {title} {Repulsion of
  resonance states and exceptional points},}\ }\href
  {https://doi.org/10.1103/physreve.61.929} {\bibfield  {journal} {\bibinfo
  {journal} {Phys. Rev. E}\ }\textbf {\bibinfo {volume} {61}},\ \bibinfo
  {pages} {929--932} (\bibinfo {year} {2000})}\BibitemShut {NoStop}%
\bibitem [{\citenamefont {Cao}\ and\ \citenamefont {Wiersig}(2015)}]{Cao2015}%
  \BibitemOpen
  \bibfield  {author} {\bibinfo {author} {\bibfnamefont {H.}~\bibnamefont
  {Cao}}\ and\ \bibinfo {author} {\bibfnamefont {J.}~\bibnamefont {Wiersig}},\
  }\bibfield  {title} {\enquote {\bibinfo {title} {Dielectric microcavities:
  Model systems for wave chaos and non-hermitian physics},}\ }\href
  {https://doi.org/10.1103/revmodphys.87.61} {\bibfield  {journal} {\bibinfo
  {journal} {Rev. Mod. Phys.}\ }\textbf {\bibinfo {volume} {87}},\ \bibinfo
  {pages} {61} (\bibinfo {year} {2015})}\BibitemShut {NoStop}%
\bibitem [{\citenamefont {Bernier}\ \emph {et~al.}(2018)\citenamefont
  {Bernier}, \citenamefont {T{\'{o}}th}, \citenamefont {Feofanov},\ and\
  \citenamefont {Kippenberg}}]{Bernier2018}%
  \BibitemOpen
  \bibfield  {author} {\bibinfo {author} {\bibfnamefont {N.~R.}\ \bibnamefont
  {Bernier}}, \bibinfo {author} {\bibfnamefont {L.~D.}\ \bibnamefont
  {T{\'{o}}th}}, \bibinfo {author} {\bibfnamefont {A.~K.}\ \bibnamefont
  {Feofanov}},\ and\ \bibinfo {author} {\bibfnamefont {T.~J.}\ \bibnamefont
  {Kippenberg}},\ }\bibfield  {title} {\enquote {\bibinfo {title} {Level
  attraction in a microwave optomechanical circuit},}\ }\href
  {https://doi.org/10.1103/physreva.98.023841} {\bibfield  {journal} {\bibinfo
  {journal} {Phys. Rev. A}\ }\textbf {\bibinfo {volume} {98}},\ \bibinfo
  {pages} {023841} (\bibinfo {year} {2018})}\BibitemShut {NoStop}%
\bibitem [{\citenamefont {Wiersig}(2006)}]{Wiersig2006}%
  \BibitemOpen
  \bibfield  {author} {\bibinfo {author} {\bibfnamefont {J.}~\bibnamefont
  {Wiersig}},\ }\bibfield  {title} {\enquote {\bibinfo {title} {Formation of
  {L}ong-{L}ived, {S}carlike {M}odes near {A}voided {R}esonance {C}rossings in
  {O}ptical {M}icrocavities},}\ }\href@noop {} {\bibfield  {journal} {\bibinfo
  {journal} {Phys. Rev. Lett.}\ }\textbf {\bibinfo {volume} {97}},\ \bibinfo
  {pages} {253901} (\bibinfo {year} {2006})}\BibitemShut {NoStop}%
\bibitem [{\citenamefont {Yang}\ \emph {et~al.}(2017)\citenamefont {Yang},
  \citenamefont {Miroshnichenko}, \citenamefont {Kostinski}, \citenamefont
  {Odit}, \citenamefont {Kapitanova}, \citenamefont {Qiu},\ and\ \citenamefont
  {Kivshar}}]{Yang2017}%
  \BibitemOpen
  \bibfield  {author} {\bibinfo {author} {\bibfnamefont {Y.}~\bibnamefont
  {Yang}}, \bibinfo {author} {\bibfnamefont {A.}~\bibnamefont
  {Miroshnichenko}}, \bibinfo {author} {\bibfnamefont {S.}~\bibnamefont
  {Kostinski}}, \bibinfo {author} {\bibfnamefont {M.}~\bibnamefont {Odit}},
  \bibinfo {author} {\bibfnamefont {P.}~\bibnamefont {Kapitanova}}, \bibinfo
  {author} {\bibfnamefont {M.}~\bibnamefont {Qiu}},\ and\ \bibinfo {author}
  {\bibfnamefont {Y.}~\bibnamefont {Kivshar}},\ }\bibfield  {title} {\enquote
  {\bibinfo {title} {Multimode directionality in all-dielectric
  metasurfaces},}\ }\href {https://doi.org/10.1103/physrevb.95.165426}
  {\bibfield  {journal} {\bibinfo  {journal} {Phys. Rev. B}\ }\textbf {\bibinfo
  {volume} {95}},\ \bibinfo {pages} {165426} (\bibinfo {year}
  {2017})}\BibitemShut {NoStop}%
\bibitem [{\citenamefont {Huang}\ \emph {et~al.}(2021)\citenamefont {Huang},
  \citenamefont {Xu}, \citenamefont {Rahmani}, \citenamefont {Neshev},\ and\
  \citenamefont {Miroshnichenko}}]{Huang2021}%
  \BibitemOpen
  \bibfield  {author} {\bibinfo {author} {\bibfnamefont {L.}~\bibnamefont
  {Huang}}, \bibinfo {author} {\bibfnamefont {L.}~\bibnamefont {Xu}}, \bibinfo
  {author} {\bibfnamefont {M.}~\bibnamefont {Rahmani}}, \bibinfo {author}
  {\bibfnamefont {D.}~\bibnamefont {Neshev}},\ and\ \bibinfo {author}
  {\bibfnamefont {A.}~\bibnamefont {Miroshnichenko}},\ }\bibfield  {title}
  {\enquote {\bibinfo {title} {Pushing the limit of high-{Q} mode of a single
  dielectric nanocavity},}\ }\href {https://doi.org/10.1117/1.ap.3.1.016004}
  {\bibfield  {journal} {\bibinfo  {journal} {Advanced Photonics}\ }\textbf
  {\bibinfo {volume} {3}},\ \bibinfo {pages} {016004} (\bibinfo {year}
  {2021})}\BibitemShut {NoStop}%
\bibitem [{\citenamefont {Song}\ and\ \citenamefont {Cao}(2010)}]{Song2010}%
  \BibitemOpen
  \bibfield  {author} {\bibinfo {author} {\bibfnamefont {Q.~H.}\ \bibnamefont
  {Song}}\ and\ \bibinfo {author} {\bibfnamefont {H.}~\bibnamefont {Cao}},\
  }\bibfield  {title} {\enquote {\bibinfo {title} {Improving optical
  confinement in nanostructures via external mode coupling},}\ }\href
  {https://doi.org/10.1103/physrevlett.105.053902} {\bibfield  {journal}
  {\bibinfo  {journal} {Phys. Rev. Lett.}\ }\textbf {\bibinfo {volume} {105}},\
  \bibinfo {pages} {053902} (\bibinfo {year} {2010})}\BibitemShut {NoStop}%
\bibitem [{\citenamefont {Rybin}\ \emph {et~al.}(2017)\citenamefont {Rybin},
  \citenamefont {Koshelev}, \citenamefont {Sadrieva}, \citenamefont {Samusev},
  \citenamefont {Bogdanov}, \citenamefont {Limonov},\ and\ \citenamefont
  {Kivshar}}]{Rybin2017}%
  \BibitemOpen
  \bibfield  {author} {\bibinfo {author} {\bibfnamefont {M.}~\bibnamefont
  {Rybin}}, \bibinfo {author} {\bibfnamefont {K.}~\bibnamefont {Koshelev}},
  \bibinfo {author} {\bibfnamefont {Z.}~\bibnamefont {Sadrieva}}, \bibinfo
  {author} {\bibfnamefont {K.}~\bibnamefont {Samusev}}, \bibinfo {author}
  {\bibfnamefont {A.}~\bibnamefont {Bogdanov}}, \bibinfo {author}
  {\bibfnamefont {M.}~\bibnamefont {Limonov}},\ and\ \bibinfo {author}
  {\bibfnamefont {Y.}~\bibnamefont {Kivshar}},\ }\bibfield  {title} {\enquote
  {\bibinfo {title} {High-{Q} {S}upercavity {M}odes in {S}ubwavelength
  {D}ielectric {R}esonators},}\ }\href
  {https://doi.org/10.1103/physrevlett.119.243901} {\bibfield  {journal}
  {\bibinfo  {journal} {Phys. Rev. Lett.}\ }\textbf {\bibinfo {volume} {119}},\
  \bibinfo {pages} {243901} (\bibinfo {year} {2017})}\BibitemShut {NoStop}%
\bibitem [{\citenamefont {Koshelev}, \citenamefont {Bogdanov},\ and\
  \citenamefont {Kivshar}(2018)}]{Koshelev2018}%
  \BibitemOpen
  \bibfield  {author} {\bibinfo {author} {\bibfnamefont {K.}~\bibnamefont
  {Koshelev}}, \bibinfo {author} {\bibfnamefont {A.}~\bibnamefont {Bogdanov}},\
  and\ \bibinfo {author} {\bibfnamefont {Y.}~\bibnamefont {Kivshar}},\
  }\bibfield  {title} {\enquote {\bibinfo {title} {Meta-optics and bound states
  in the continuum},}\ }\href@noop {} {\bibfield  {journal} {\bibinfo
  {journal} {Science Bulletin}\ }\textbf {\bibinfo {volume} {17}},\ \bibinfo
  {pages} {065601} (\bibinfo {year} {2018})}\BibitemShut {NoStop}%
\bibitem [{\citenamefont {Chen}, \citenamefont {Chen},\ and\ \citenamefont
  {Liu}(2019)}]{Chen2019}%
  \BibitemOpen
  \bibfield  {author} {\bibinfo {author} {\bibfnamefont {W.}~\bibnamefont
  {Chen}}, \bibinfo {author} {\bibfnamefont {Y.}~\bibnamefont {Chen}},\ and\
  \bibinfo {author} {\bibfnamefont {W.}~\bibnamefont {Liu}},\ }\bibfield
  {title} {\enquote {\bibinfo {title} {Multipolar conversion induced
  subwavelength high-q kerker supermodes with unidirectional radiations},}\
  }\href {https://doi.org/10.1002/lpor.201900067} {\bibfield  {journal}
  {\bibinfo  {journal} {Laser {\&} Photonics Reviews}\ }\textbf {\bibinfo
  {volume} {13}},\ \bibinfo {pages} {1900067} (\bibinfo {year}
  {2019})}\BibitemShut {NoStop}%
\bibitem [{\citenamefont {Wang}\ \emph {et~al.}(2019)\citenamefont {Wang},
  \citenamefont {Zheng}, \citenamefont {Xiong}, \citenamefont {Qi},\ and\
  \citenamefont {Li}}]{Wang2019}%
  \BibitemOpen
  \bibfield  {author} {\bibinfo {author} {\bibfnamefont {W.}~\bibnamefont
  {Wang}}, \bibinfo {author} {\bibfnamefont {L.}~\bibnamefont {Zheng}},
  \bibinfo {author} {\bibfnamefont {L.}~\bibnamefont {Xiong}}, \bibinfo
  {author} {\bibfnamefont {J.}~\bibnamefont {Qi}},\ and\ \bibinfo {author}
  {\bibfnamefont {B.}~\bibnamefont {Li}},\ }\bibfield  {title} {\enquote
  {\bibinfo {title} {High {Q}-factor multiple fano resonances for
  high-sensitivity sensing in all-dielectric metamaterials},}\ }\href
  {https://doi.org/10.1364/osac.2.002818} {\bibfield  {journal} {\bibinfo
  {journal} {{OSA} Continuum}\ }\textbf {\bibinfo {volume} {2}},\ \bibinfo
  {pages} {2818} (\bibinfo {year} {2019})}\BibitemShut {NoStop}%
\bibitem [{\citenamefont {Odit}\ \emph {et~al.}(2020)\citenamefont {Odit},
  \citenamefont {Koshelev}, \citenamefont {Gladyshev}, \citenamefont
  {Ladutenko}, \citenamefont {Kivshar},\ and\ \citenamefont
  {Bogdanov}}]{Odit2020}%
  \BibitemOpen
  \bibfield  {author} {\bibinfo {author} {\bibfnamefont {M.}~\bibnamefont
  {Odit}}, \bibinfo {author} {\bibfnamefont {K.}~\bibnamefont {Koshelev}},
  \bibinfo {author} {\bibfnamefont {S.}~\bibnamefont {Gladyshev}}, \bibinfo
  {author} {\bibfnamefont {K.}~\bibnamefont {Ladutenko}}, \bibinfo {author}
  {\bibfnamefont {Y.}~\bibnamefont {Kivshar}},\ and\ \bibinfo {author}
  {\bibfnamefont {A.}~\bibnamefont {Bogdanov}},\ }\bibfield  {title} {\enquote
  {\bibinfo {title} {Observation of supercavity modes in subwavelength
  dielectric resonators},}\ }\href {https://doi.org/10.1002/adma.202003804}
  {\bibfield  {journal} {\bibinfo  {journal} {Advanced Materials}\ ,\ \bibinfo
  {pages} {2003804}} (\bibinfo {year} {2020})}\BibitemShut {NoStop}%
\bibitem [{\citenamefont {Volkovskaya}\ \emph {et~al.}(2020)\citenamefont
  {Volkovskaya}, \citenamefont {Xu}, \citenamefont {Huang}, \citenamefont
  {Smirnov}, \citenamefont {Miroshnichenko},\ and\ \citenamefont
  {Smirnova}}]{Volkovskaya2020}%
  \BibitemOpen
  \bibfield  {author} {\bibinfo {author} {\bibfnamefont {I.}~\bibnamefont
  {Volkovskaya}}, \bibinfo {author} {\bibfnamefont {L.}~\bibnamefont {Xu}},
  \bibinfo {author} {\bibfnamefont {L.}~\bibnamefont {Huang}}, \bibinfo
  {author} {\bibfnamefont {A.}~\bibnamefont {Smirnov}}, \bibinfo {author}
  {\bibfnamefont {A.}~\bibnamefont {Miroshnichenko}},\ and\ \bibinfo {author}
  {\bibfnamefont {D.}~\bibnamefont {Smirnova}},\ }\bibfield  {title} {\enquote
  {\bibinfo {title} {Multipolar second-harmonic generation from high-q
  quasi-{BIC} states in subwavelength resonators},}\ }\href
  {https://doi.org/10.1515/nanoph-2020-0156} {\bibfield  {journal} {\bibinfo
  {journal} {Nanophotonics}\ }\textbf {\bibinfo {volume} {9}},\ \bibinfo
  {pages} {3953} (\bibinfo {year} {2020})}\BibitemShut {NoStop}%
\bibitem [{\citenamefont {Boriskina}(2006)}]{Boriskina2006}%
  \BibitemOpen
  \bibfield  {author} {\bibinfo {author} {\bibfnamefont {S.~V.}\ \bibnamefont
  {Boriskina}},\ }\bibfield  {title} {\enquote {\bibinfo {title} {Theoretical
  prediction of a dramatic q-factor enhancement and degeneracy removal of
  whispering gallery modes in symmetrical photonic molecules},}\ }\href
  {https://doi.org/10.1364/ol.31.000338} {\bibfield  {journal} {\bibinfo
  {journal} {Opt. Lett.}\ }\textbf {\bibinfo {volume} {31}},\ \bibinfo {pages}
  {338} (\bibinfo {year} {2006})}\BibitemShut {NoStop}%
\bibitem [{\citenamefont {Boriskina}(2007)}]{Boriskina2007}%
  \BibitemOpen
  \bibfield  {author} {\bibinfo {author} {\bibfnamefont {S.}~\bibnamefont
  {Boriskina}},\ }\bibfield  {title} {\enquote {\bibinfo {title} {Coupling of
  whispering-gallery modes in size-mismatched microdisk photonic molecules},}\
  }\href {https://doi.org/10.1364/ol.32.001557} {\bibfield  {journal} {\bibinfo
   {journal} {Opt. Lett.}\ }\textbf {\bibinfo {volume} {32}},\ \bibinfo {pages}
  {1557} (\bibinfo {year} {2007})}\BibitemShut {NoStop}%
\bibitem [{\citenamefont {Benyoucef}\ \emph {et~al.}(2011)\citenamefont
  {Benyoucef}, \citenamefont {Shim}, \citenamefont {Wiersig},\ and\
  \citenamefont {Schmidt}}]{Benyoucef2011}%
  \BibitemOpen
  \bibfield  {author} {\bibinfo {author} {\bibfnamefont {M.}~\bibnamefont
  {Benyoucef}}, \bibinfo {author} {\bibfnamefont {J.-B.}\ \bibnamefont {Shim}},
  \bibinfo {author} {\bibfnamefont {J.}~\bibnamefont {Wiersig}},\ and\ \bibinfo
  {author} {\bibfnamefont {O.~G.}\ \bibnamefont {Schmidt}},\ }\bibfield
  {title} {\enquote {\bibinfo {title} {Quality-factor enhancement of supermodes
  in coupled microdisks},}\ }\href {https://doi.org/10.1364/ol.36.001317}
  {\bibfield  {journal} {\bibinfo  {journal} {Opt. Lett.}\ }\textbf {\bibinfo
  {volume} {36}},\ \bibinfo {pages} {1317} (\bibinfo {year}
  {2011})}\BibitemShut {NoStop}%
\bibitem [{\citenamefont {Chang}\ and\ \citenamefont {Sun}(2016)}]{Chang2016}%
  \BibitemOpen
  \bibfield  {author} {\bibinfo {author} {\bibfnamefont {S.-H.~G.}\
  \bibnamefont {Chang}}\ and\ \bibinfo {author} {\bibfnamefont {C.-Y.}\
  \bibnamefont {Sun}},\ }\bibfield  {title} {\enquote {\bibinfo {title}
  {Avoided resonance crossing and non-reciprocal nearly perfect absorption in
  plasmonic nanodisks with near-field and far-field couplings},}\ }\href
  {https://doi.org/10.1364/oe.24.016822} {\bibfield  {journal} {\bibinfo
  {journal} {Optics Express}\ }\textbf {\bibinfo {volume} {24}},\ \bibinfo
  {pages} {16822} (\bibinfo {year} {2016})}\BibitemShut {NoStop}%
\bibitem [{\citenamefont {Pichugin}\ and\ \citenamefont
  {Sadreev}(2019)}]{Pichugin2019}%
  \BibitemOpen
  \bibfield  {author} {\bibinfo {author} {\bibfnamefont {K.~N.}\ \bibnamefont
  {Pichugin}}\ and\ \bibinfo {author} {\bibfnamefont {A.~F.}\ \bibnamefont
  {Sadreev}},\ }\bibfield  {title} {\enquote {\bibinfo {title} {Interaction
  between coaxial dielectric disks enhances the {Q} factor},}\ }\href
  {https://doi.org/10.1063/1.5094188} {\bibfield  {journal} {\bibinfo
  {journal} {J. Appl. Phys.}\ }\textbf {\bibinfo {volume} {126}},\ \bibinfo
  {pages} {093105} (\bibinfo {year} {2019})}\BibitemShut {NoStop}%
\bibitem [{\citenamefont {Bulgakov}, \citenamefont {Pichugin},\ and\
  \citenamefont {Sadreev}(2019)}]{Bulgakov2019}%
  \BibitemOpen
  \bibfield  {author} {\bibinfo {author} {\bibfnamefont {E.~N.}\ \bibnamefont
  {Bulgakov}}, \bibinfo {author} {\bibfnamefont {K.~N.}\ \bibnamefont
  {Pichugin}},\ and\ \bibinfo {author} {\bibfnamefont {A.~F.}\ \bibnamefont
  {Sadreev}},\ }\bibfield  {title} {\enquote {\bibinfo {title} {Evolution of
  the resonances of two parallel dielectric cylinders with distance between
  them},}\ }\href {https://doi.org/10.1103/physreva.100.043806} {\bibfield
  {journal} {\bibinfo  {journal} {Phys. Rev. A}\ }\textbf {\bibinfo {volume}
  {100}},\ \bibinfo {pages} {043806} (\bibinfo {year} {2019})}\BibitemShut
  {NoStop}%
\bibitem [{\citenamefont {Dmitriev}\ and\ \citenamefont
  {Rybin}(2019)}]{Dmitriev2019}%
  \BibitemOpen
  \bibfield  {author} {\bibinfo {author} {\bibfnamefont {A.}~\bibnamefont
  {Dmitriev}}\ and\ \bibinfo {author} {\bibfnamefont {M.}~\bibnamefont
  {Rybin}},\ }\bibfield  {title} {\enquote {\bibinfo {title} {Combining
  isolated scatterers into a dimer by strong optical coupling},}\ }\href
  {https://doi.org/10.1103/physreva.99.063837} {\bibfield  {journal} {\bibinfo
  {journal} {Phys. Rev. A}\ }\textbf {\bibinfo {volume} {99}},\ \bibinfo
  {pages} {063837} (\bibinfo {year} {2019})}\BibitemShut {NoStop}%
\bibitem [{\citenamefont {Vinel}\ \emph {et~al.}(2021)\citenamefont {Vinel},
  \citenamefont {Li}, \citenamefont {Borne}, \citenamefont {Bensemhoun},
  \citenamefont {Favero}, \citenamefont {Ciuti},\ and\ \citenamefont
  {Leo}}]{Vinel2021}%
  \BibitemOpen
  \bibfield  {author} {\bibinfo {author} {\bibfnamefont {V.}~\bibnamefont
  {Vinel}}, \bibinfo {author} {\bibfnamefont {Z.}~\bibnamefont {Li}}, \bibinfo
  {author} {\bibfnamefont {A.}~\bibnamefont {Borne}}, \bibinfo {author}
  {\bibfnamefont {A.}~\bibnamefont {Bensemhoun}}, \bibinfo {author}
  {\bibfnamefont {I.}~\bibnamefont {Favero}}, \bibinfo {author} {\bibfnamefont
  {C.}~\bibnamefont {Ciuti}},\ and\ \bibinfo {author} {\bibfnamefont
  {G.}~\bibnamefont {Leo}},\ }\bibfield  {title} {\enquote {\bibinfo {title}
  {Non-hermitian bath model for arrays of coupled nanoresonators},}\ }\href
  {https://doi.org/10.1364/oe.436125} {\bibfield  {journal} {\bibinfo
  {journal} {Optics Express}\ }\textbf {\bibinfo {volume} {29}},\ \bibinfo
  {pages} {34015} (\bibinfo {year} {2021})}\BibitemShut {NoStop}%
\bibitem [{\citenamefont {Pichugin}, \citenamefont {Sadreev},\ and\
  \citenamefont {Bulgakov}(2021)}]{Pichugin2021}%
  \BibitemOpen
  \bibfield  {author} {\bibinfo {author} {\bibfnamefont {K.}~\bibnamefont
  {Pichugin}}, \bibinfo {author} {\bibfnamefont {A.}~\bibnamefont {Sadreev}},\
  and\ \bibinfo {author} {\bibfnamefont {E.}~\bibnamefont {Bulgakov}},\
  }\bibfield  {title} {\enquote {\bibinfo {title} {Ultrahigh-{Q} system of a
  few coaxial disks},}\ }\href {https://doi.org/10.1515/nanoph-2021-0345}
  {\bibfield  {journal} {\bibinfo  {journal} {Nanophotonics}\ }\textbf
  {\bibinfo {volume} {10}},\ \bibinfo {pages} {4341--4346} (\bibinfo {year}
  {2021})}\BibitemShut {NoStop}%
\bibitem [{\citenamefont {Kitamura}, \citenamefont {Pilon},\ and\ \citenamefont
  {Jonasz}(2007)}]{Kitamura2007}%
  \BibitemOpen
  \bibfield  {author} {\bibinfo {author} {\bibfnamefont {R.}~\bibnamefont
  {Kitamura}}, \bibinfo {author} {\bibfnamefont {L.}~\bibnamefont {Pilon}},\
  and\ \bibinfo {author} {\bibfnamefont {M.}~\bibnamefont {Jonasz}},\
  }\bibfield  {title} {\enquote {\bibinfo {title} {Optical constants of silica
  glass from extreme ultraviolet to far infrared at near room temperature},}\
  }\href {https://doi.org/10.1364/ao.46.008118} {\bibfield  {journal} {\bibinfo
   {journal} {Appl. Optics}\ }\textbf {\bibinfo {volume} {46}},\ \bibinfo
  {pages} {8118} (\bibinfo {year} {2007})}\BibitemShut {NoStop}%
\bibitem [{\citenamefont {Bulgakov}, \citenamefont {Pichugin},\ and\
  \citenamefont {Sadreev}(2021{\natexlab{a}})}]{Bulgakov2021}%
  \BibitemOpen
  \bibfield  {author} {\bibinfo {author} {\bibfnamefont {E.}~\bibnamefont
  {Bulgakov}}, \bibinfo {author} {\bibfnamefont {K.}~\bibnamefont {Pichugin}},\
  and\ \bibinfo {author} {\bibfnamefont {A.}~\bibnamefont {Sadreev}},\
  }\bibfield  {title} {\enquote {\bibinfo {title} {Mie resonance engineering in
  two disks},}\ }\href {https://doi.org/10.3390/photonics8020049} {\bibfield
  {journal} {\bibinfo  {journal} {MDPI Photonics}\ }\textbf {\bibinfo {volume}
  {8}},\ \bibinfo {pages} {49} (\bibinfo {year}
  {2021}{\natexlab{a}})}\BibitemShut {NoStop}%
\bibitem [{\citenamefont {Jackson}(1962)}]{Jackson}%
  \BibitemOpen
  \bibfield  {author} {\bibinfo {author} {\bibfnamefont {J.~D.}\ \bibnamefont
  {Jackson}},\ }\href@noop {} {\emph {\bibinfo {title} {Classical
  Electrodynamics}}}\ (\bibinfo  {publisher} {John Wiley and Sons, Inc.,},\
  \bibinfo {address} {New York},\ \bibinfo {year} {1962})\BibitemShut {NoStop}%
\bibitem [{\citenamefont {Bogdanov}\ \emph {et~al.}(2019)\citenamefont
  {Bogdanov}, \citenamefont {Koshelev}, \citenamefont {Kapitanova},
  \citenamefont {Rybin}, \citenamefont {Gladyshev}, \citenamefont {Sadrieva},
  \citenamefont {Samusev}, \citenamefont {Kivshar},\ and\ \citenamefont
  {Limonov}}]{Bogdanov2019}%
  \BibitemOpen
  \bibfield  {author} {\bibinfo {author} {\bibfnamefont {A.}~\bibnamefont
  {Bogdanov}}, \bibinfo {author} {\bibfnamefont {K.}~\bibnamefont {Koshelev}},
  \bibinfo {author} {\bibfnamefont {P.}~\bibnamefont {Kapitanova}}, \bibinfo
  {author} {\bibfnamefont {M.}~\bibnamefont {Rybin}}, \bibinfo {author}
  {\bibfnamefont {S.}~\bibnamefont {Gladyshev}}, \bibinfo {author}
  {\bibfnamefont {Z.}~\bibnamefont {Sadrieva}}, \bibinfo {author}
  {\bibfnamefont {K.}~\bibnamefont {Samusev}}, \bibinfo {author} {\bibfnamefont
  {Y.}~\bibnamefont {Kivshar}},\ and\ \bibinfo {author} {\bibfnamefont {M.~F.}\
  \bibnamefont {Limonov}},\ }\bibfield  {title} {\enquote {\bibinfo {title}
  {Bound states in the continuum and {F}ano resonances in the strong mode
  coupling regime},}\ }\href {https://doi.org/10.1117/1.ap.1.1.016001}
  {\bibfield  {journal} {\bibinfo  {journal} {Adv. Photonics}\ }\textbf
  {\bibinfo {volume} {1}},\ \bibinfo {pages} {1} (\bibinfo {year}
  {2019})}\BibitemShut {NoStop}%
\bibitem [{\citenamefont {Sadrieva}\ \emph
  {et~al.}(2019{\natexlab{b}})\citenamefont {Sadrieva}, \citenamefont
  {Frizyuk}, \citenamefont {Petrov}, \citenamefont {Kivshar},\ and\
  \citenamefont {Bogdanov}}]{Sadrieva2019a}%
  \BibitemOpen
  \bibfield  {author} {\bibinfo {author} {\bibfnamefont {Z.}~\bibnamefont
  {Sadrieva}}, \bibinfo {author} {\bibfnamefont {K.}~\bibnamefont {Frizyuk}},
  \bibinfo {author} {\bibfnamefont {M.}~\bibnamefont {Petrov}}, \bibinfo
  {author} {\bibfnamefont {Y.}~\bibnamefont {Kivshar}},\ and\ \bibinfo {author}
  {\bibfnamefont {A.}~\bibnamefont {Bogdanov}},\ }\bibfield  {title} {\enquote
  {\bibinfo {title} {Multipolar origin of bound states in the continuum},}\
  }\href {https://doi.org/10.1103/physrevb.100.115303} {\bibfield  {journal}
  {\bibinfo  {journal} {Phys. Rev. B}\ }\textbf {\bibinfo {volume} {100}}
  (\bibinfo {year} {2019}{\natexlab{b}}),\
  10.1103/physrevb.100.115303}\BibitemShut {NoStop}%
\bibitem [{\citenamefont {Stratton}(1941)}]{Stratton}%
  \BibitemOpen
  \bibfield  {author} {\bibinfo {author} {\bibfnamefont {J.~A.}\ \bibnamefont
  {Stratton}},\ }\href@noop {} {\emph {\bibinfo {title} {Electromagnetic
  theory}}},\ edited by\ \bibinfo {editor} {\bibfnamefont {L.~A.}\ \bibnamefont
  {DuBridge}}\ (\bibinfo  {publisher} {McGraw-Hill Book Company, Inc.},\
  \bibinfo {year} {1941})\BibitemShut {NoStop}%
\bibitem [{\citenamefont {Linton}, \citenamefont {Zalipaev},\ and\
  \citenamefont {Thompson}(2013)}]{Linton2013}%
  \BibitemOpen
  \bibfield  {author} {\bibinfo {author} {\bibfnamefont {C.}~\bibnamefont
  {Linton}}, \bibinfo {author} {\bibfnamefont {V.}~\bibnamefont {Zalipaev}},\
  and\ \bibinfo {author} {\bibfnamefont {I.}~\bibnamefont {Thompson}},\
  }\bibfield  {title} {\enquote {\bibinfo {title} {Electromagnetic guided waves
  on linear arrays of spheres},}\ }\href
  {https://doi.org/10.1016/j.wavemoti.2012.06.002} {\bibfield  {journal}
  {\bibinfo  {journal} {Wave Motion}\ }\textbf {\bibinfo {volume} {50}},\
  \bibinfo {pages} {29--40} (\bibinfo {year} {2013})}\BibitemShut {NoStop}%
\bibitem [{\citenamefont {Bulgakov}, \citenamefont {Pichugin},\ and\
  \citenamefont {Sadreev}(2021{\natexlab{b}})}]{Bulgakov2021b}%
  \BibitemOpen
  \bibfield  {author} {\bibinfo {author} {\bibfnamefont {E.}~\bibnamefont
  {Bulgakov}}, \bibinfo {author} {\bibfnamefont {K.}~\bibnamefont {Pichugin}},\
  and\ \bibinfo {author} {\bibfnamefont {A.}~\bibnamefont {Sadreev}},\
  }\bibfield  {title} {\enquote {\bibinfo {title} {Exceptional points in a
  dielectric spheroid},}\ }\href {https://doi.org/10.1103/physreva.104.053507}
  {\bibfield  {journal} {\bibinfo  {journal} {Phys. Rev. A}\ }\textbf {\bibinfo
  {volume} {104}},\ \bibinfo {pages} {053507} (\bibinfo {year}
  {2021}{\natexlab{b}})}\BibitemShut {NoStop}%
\bibitem [{\citenamefont {Doicu}, \citenamefont {Wriedt},\ and\ \citenamefont
  {Eremin}(2006)}]{Doicu2006}%
  \BibitemOpen
  \bibfield  {author} {\bibinfo {author} {\bibfnamefont {A.}~\bibnamefont
  {Doicu}}, \bibinfo {author} {\bibfnamefont {T.}~\bibnamefont {Wriedt}},\ and\
  \bibinfo {author} {\bibfnamefont {Y.~A.}\ \bibnamefont {Eremin}},\ }\href
  {https://doi.org/10.1007/978-3-540-33697-6} {\emph {\bibinfo {title} {Light
  Scattering by Systems of Particles}}}\ (\bibinfo  {publisher} {Springer
  Berlin Heidelberg},\ \bibinfo {year} {2006})\BibitemShut {NoStop}%
\bibitem [{\citenamefont {Hu}\ and\ \citenamefont {Weiss}(2016)}]{Hu2016}%
  \BibitemOpen
  \bibfield  {author} {\bibinfo {author} {\bibfnamefont {S.}~\bibnamefont
  {Hu}}\ and\ \bibinfo {author} {\bibfnamefont {S.}~\bibnamefont {Weiss}},\
  }\bibfield  {title} {\enquote {\bibinfo {title} {Design of photonic crystal
  cavities for extreme light concentration},}\ }\href
  {https://doi.org/10.1021/acsphotonics.6b00219} {\bibfield  {journal}
  {\bibinfo  {journal} {{ACS} Photonics}\ }\textbf {\bibinfo {volume} {3}},\
  \bibinfo {pages} {1647} (\bibinfo {year} {2016})}\BibitemShut {NoStop}%
\bibitem [{\citenamefont {Vennberg}, \citenamefont {Ravishankar},\ and\
  \citenamefont {Anand.}(2022)}]{Vennberg2022}%
  \BibitemOpen
  \bibfield  {author} {\bibinfo {author} {\bibfnamefont {F.}~\bibnamefont
  {Vennberg}}, \bibinfo {author} {\bibfnamefont {A.}~\bibnamefont
  {Ravishankar}},\ and\ \bibinfo {author} {\bibfnamefont {S.}~\bibnamefont
  {Anand.}},\ }\bibfield  {title} {\enquote {\bibinfo {title} {Manipulating
  light scattering and optical confinement in vertically stacked {M}ie
  resonators},}\ }\href {https://doi.org/10.1515/nanoph-2022-0605} {\bibfield
  {journal} {\bibinfo  {journal} {Nanophotonics}\ }\textbf {\bibinfo {volume}
  {11}},\ \bibinfo {pages} {4755} (\bibinfo {year} {2022})}\BibitemShut
  {NoStop}%
\bibitem [{\citenamefont {Maslova}\ \emph {et~al.}(2021)\citenamefont
  {Maslova}, \citenamefont {Rybin}, \citenamefont {Bogdanov},\ and\
  \citenamefont {Sadrieva}}]{Maslova2021}%
  \BibitemOpen
  \bibfield  {author} {\bibinfo {author} {\bibfnamefont {E.}~\bibnamefont
  {Maslova}}, \bibinfo {author} {\bibfnamefont {M.}~\bibnamefont {Rybin}},
  \bibinfo {author} {\bibfnamefont {A.}~\bibnamefont {Bogdanov}},\ and\
  \bibinfo {author} {\bibfnamefont {Z.}~\bibnamefont {Sadrieva}},\ }\bibfield
  {title} {\enquote {\bibinfo {title} {Bound states in the continuum in
  periodic structures with structural disorder},}\ }\href
  {https://doi.org/10.1515/nanoph-2021-0475} {\bibfield  {journal} {\bibinfo
  {journal} {Nanophotonics}\ }\textbf {\bibinfo {volume} {10}},\ \bibinfo
  {pages} {4313--4321} (\bibinfo {year} {2021})}\BibitemShut {NoStop}%
\bibitem [{\citenamefont {Kim}, \citenamefont {Kim},\ and\ \citenamefont
  {Cahoon}(2019)}]{Kim2019}%
  \BibitemOpen
  \bibfield  {author} {\bibinfo {author} {\bibfnamefont {S.}~\bibnamefont
  {Kim}}, \bibinfo {author} {\bibfnamefont {K.-H.}\ \bibnamefont {Kim}},\ and\
  \bibinfo {author} {\bibfnamefont {J.}~\bibnamefont {Cahoon}},\ }\bibfield
  {title} {\enquote {\bibinfo {title} {Optical bound states in the continuum
  with nanowire geometric superlattices},}\ }\href
  {https://doi.org/10.1103/physrevlett.122.187402} {\bibfield  {journal}
  {\bibinfo  {journal} {Phys. Rev. Lett.}\ }\textbf {\bibinfo {volume} {122}},\
  \bibinfo {pages} {187402} (\bibinfo {year} {2019})}\BibitemShut {NoStop}%
\end{thebibliography}
%

\end{document}